\documentclass{aa}

\usepackage{graphicx}
\usepackage{longtable,rotating}
\usepackage{booktabs}
\usepackage{amsmath}
\usepackage{natbib}
\bibpunct{(}{)}{;}{a}{}{,} 
\usepackage{placeins}
\usepackage{color}
\usepackage[x11names]{xcolor}
\usepackage{arydshln}
\usepackage{hyperref}
\hypersetup{linkcolor= blue, citecolor=blue, colorlinks=True, pdfborder={0 0 0} }
\usepackage{txfonts}
\usepackage{pdflscape} 
\usepackage{float}
\usepackage{morefloats}
\usepackage{soul}

\newcommand{\msun}{\mbox{M$_\odot$}}

\begin{document}

\title{
The \textit{Herschel} view of the dense core population in the Ophiuchus molecular cloud
}

\author{
B. Ladjelate\inst{1,2} 
\and 
Ph. Andr\'e\inst{2} 
\and 
V. K\"onyves\inst{2,3} 
\and 
D.~Ward-Thompson\inst{3}
\and 
A. Men'shchikov\inst{2} 
\and 
A. Bracco\inst{4} 
\and 
P. Palmeirim\inst{5} 
\and 
A. Roy\inst{2} 
\and 
Y. Shimajiri\inst{6,7} 
\and
J.M. Kirk\inst{3}
\and
D. Arzoumanian\inst{5}
\and
M. Benedettini\inst{8}
\and
J. Di Francesco\inst{9}
\and 
E. Fiorellino\inst{12,13,14}
\and
N. Schneider\inst{10,11}
\and	 
S. Pezzuto\inst{7}
\and
the Herschel Gould Belt Survey Team 
}

\institute{
Instituto de Radioastronomía Milimétrica, IRAM Avenida Divina Pastora 7, Local 20, 18012, Granada, Spain
\and
Laboratoire d'Astrophysique (AIM), CEA/DRF, CNRS, Universit\'e Paris-Saclay, Universit\'e Paris Diderot, Sorbonne Paris Cit\'e, 91191 Gif-sur-Yvette, France \\
\email{ladjelate@iram.es}
\and
Jeremiah Horrocks Institute, University of Central Lancashire, Preston, Lancashire, PR1 2HE, UK
\and
Laboratoire de Physique de l’Ecole Normale Supérieure, ENS, Université PSL, CNRS, Sorbonne Université, Université de Paris, Paris, France
\and
Instituto de Astrof\'isica e Ci{\^e}ncias do Espa\c{c}o, Universidade do Porto, CAUP, Rua das Estrelas, PT4150-762 Porto, Portugal
\and
Department of Physics and Astronomy, Graduate School of Science and Engineering, Kagoshima University, 1-21-35 Korimoto, Kagoshima, Kagoshima 890-0065, Japan
\and
National Astronomical Observatory of Japan, Osawa 2-21-1, Mitaka, Tokyo 181-8588, Japan
\and
INAF - Istituto di Astrofisica e Planetologia Spaziali, Via Fosso del
Cavaliere 100, I-00133 Roma, Italy
\and
National Research Council of Canada, Herzberg Astronomy \& Astrophysics Research Centre, 5071 West Saanich Road Victoria, BC V9E 2E7, Canada
\and
I. Physik. Institut, University of Cologne, 50937, Cologne, Germany
\and
Laboratoire d'Astrophysique de Bordeaux, CNRS/INSU, Université de Bordeaux, UMR, 5804, France
\and
Dipartimento di Fisica, Università di Roma `Tor Vergata' Via della Ricerca Scientifica 1, 00133, Roma, Italy
\and 
INAF-Osservatorio Astronomico di Roma, via di Frascati 33, 00078, Monte Porzio Catone, Italy
\and 
European Southern Observatory, Karl-Schwarzschild-Strasse 2, 85748 Garching bei München, Germany
}

  \abstract
    {{\it Herschel} observations of nearby clouds in the Gould Belt support a paradigm for low-mass star formation, starting with the generation of molecular filaments, followed by filament fragmentation, and the concentration of mass into self-gravitating prestellar cores. 
    In the case of the Ophiuchus molecular complex, a rich star formation activity has been documented for many years inside the clumps of L1688, the main and densest cloud of the complex, and in the more quiescent twin cloud L1689 thanks to extensive surveys at infrared and other wavelengths.}
    {With the unique far-infrared and submillimeter continuum imaging capabilities of the {\it Herschel} Space observatory, the closeby ($d = 139\,$pc) Ophiuchus cloud was extensively mapped at five wavelengths from 70\,$\mu$m to 500\,$\mu$m with the aim of providing a complete census of dense cores in this region, including unbound starless cores, bound prestellar cores, and protostellar cores.} 
    {Taking full advantage of the high dynamic range and multi-wavelength nature of the  {\it Herschel} data, we used the multi-scale decomposition algorithms \textsl{getsources} and \textsl{getfilaments} 
    to identify an essentially complete sample of dense cores and filaments in the cloud and study their properties.}
    {The densest clouds of the Ophiuchus complex, L1688 and L1689, which thus far are only indirectly described as filamentary regions owing to the spatial distribution of their young stellar objects (YSOs), are now confirmed to be dominated by filamentary structures. The tight correlation observed between prestellar cores  and filamentary structures in L1688 and L1689 supports the view that solar-type star formation occurs primarily in dense filaments. 
    While the sub clouds of the complex show some disparities, L1689 being apparently less efficient than L1688 at forming stars when considering their total mass budgets,  both sub clouds share almost the same prestellar core formation efficiency in dense molecular gas. We also find evidence in the {\it Herschel} data for  a remarkable concentric geometrical configuration in L1688 which is dominated by up to three arc-like compression fronts and presumably created by shockwave events emanating from the Sco~OB2 association, including the neighboring massive (O9V) star $\sigma$\,Sco.}
    {Our {\it Herschel} study of the well-documented Ophiuchus region has allowed us to further analyze the influence of several early-type (OB) stars surrounding the complex, thus providing positive feedback and enhancing star formation activity in the dense central part of the region, L1688.}


\titlerunning{\emph{Herschel} Gould Belt survey results in Ophiuchus}

\maketitle

\section{Introduction}
Our general observational understanding of star formation in the Milky Way has greatly advanced in the past decade with the exploitation of extensive surveys at infrared and submillimeter wavelengths with the {\it Spitzer} and {\it Herschel}  space observatories (e.g., \citealp{2009ApJS..181..321E}, \citealp{2010A&A...518L.102A}, \citealp{2010A&A...518L.100M}, \citealp{2015ApJS..220...11D}). 
In particular, {\it Herschel} has made way for a new horizon in the observational characterization of the density structure of molecular clouds thanks to its unique imaging capabilities in the far-infrared and submillimeter regime, which probe the structure of cold cloud material down to low column densities and unveil some aspects of the star formation process at relatively high resolution inside the most deeply embedded areas of molecular clouds.

Prestellar cores represent the initial stages of low-mass star formation (see \citealp{2007prpl.conf...17D}, \citealp{2007prpl.conf...33W}) and studying their connection with the properties and structure of the parent molecular clouds is crucial in understanding the processes leading to populations and clusters of young stars.
The {\it Herschel} Gould Belt Survey (HGBS - \citealp{2010A&A...518L.102A}) has provided a significant step forward in our understanding of the link between cloud structure and the formation of prestellar cores.
This survey of nearby molecular clouds, at distances between $\sim 130\,$pc and $\sim 500\,$pc from the Sun, led to important first-look results, which were then confirmed by more extensive  studies in Aquila, Taurus, and Corona Australis for example (\citealp{2015A&A...584A..91K}, \citealp{2016MNRAS.459..342M}, \citealp{2018A&A...615A.125B}, \citealp{2018A&A...619A..52B}). 
Overall, HGBS studies confirm the ubiquity of $\sim\, $0.1\,pc-wide filaments in nearby molecular clouds \citep[][]{Arzoumanian+2011, Arzoumanian+2019} and highlight the key role of filaments in the core and star formation process (\citealp{2014prpl.conf...27A}).

Here, we report the results of the HGBS census of dense cores and molecular filaments in the Ophiuchus cloud complex (L1688/L1689). 
The outline of the paper is as follows. 
Section~2 introduces the region and summarizes the results of previous observations on the Ophiuchus molecular cloud. 
In Sect.~3, we describe the {\it Herschel} observations as well as the data reduction performed to provide scientific images. 
Section~4 is dedicated to a complete description of the {\it Herschel} GBS map products, including dust temperature and column-density maps. 
This section also provides details on the procedure adopted to obtain a census of filaments and to extract and characterize the dense cores of the region. 
In Sect.~5, the properties of the Ophiuchus cloud complex and its objects are discussed, including the close relationship between prestellar cores and filaments. 
A star formation scenario relying on the external influence of various early-type (OB) stars in the surrounding area is also explored.
Section~6 concludes the paper.

\section{The Ophiuchus molecular cloud}
\begin{figure*}[!h]
\includegraphics[trim={0cm 0cm 0cm 0.3cm},clip,width=0.99\hsize]{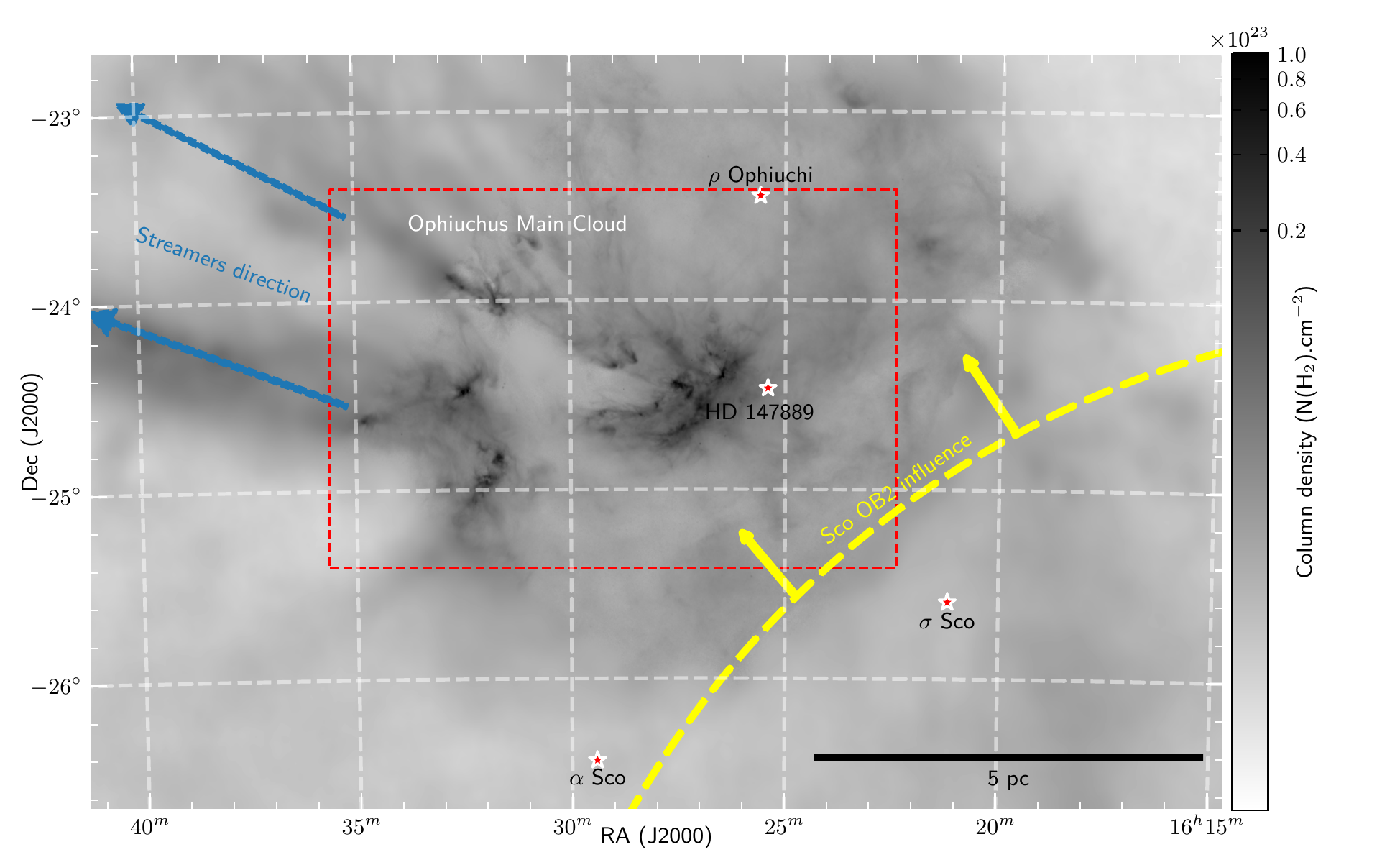}
\includegraphics[width=0.99\hsize]{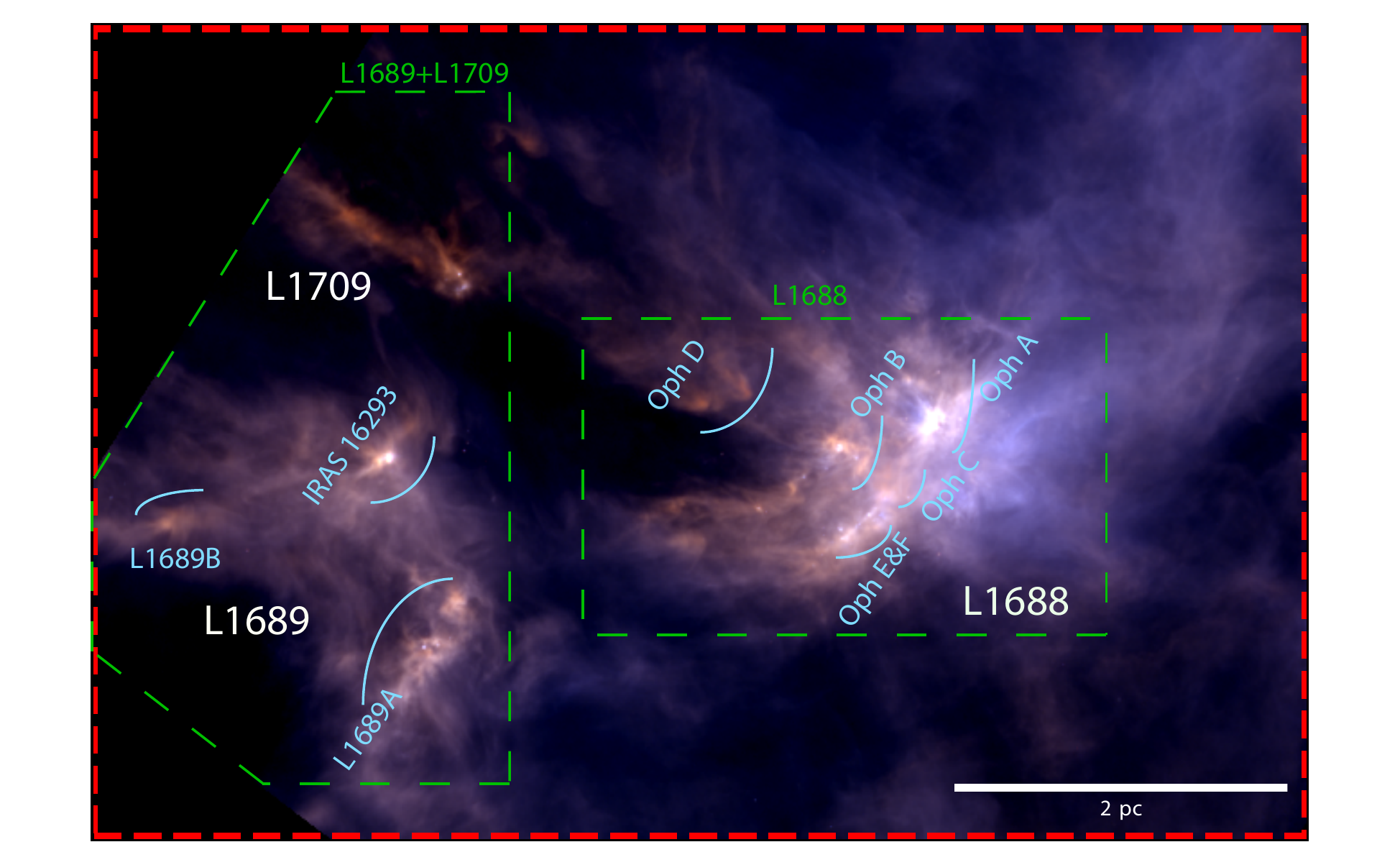}
\caption{{\bf (Top)} Multi-resolution column density map of the Ophiuchus complex derived 
from {\it Planck}/{\it Herschel} data (see text). The red rectangle shows the area encompassing the Ophiuchus main cloud covered by the {\it Herschel} data discussed in the paper.
The effective resolution ranges from $18.2\arcsec$ in this area to $5\arcmin $ in the outer parts.
{\bf (Bottom)} Composite three-color image of the field outlined by the red rectangle in the top panel,  combining {\it Herschel}/PACS 160 $\mu$m data as a  blue layer, {\it Herschel}/SPIRE 250 $\mu$m data as a green layer, and {\it Herschel}/SPIRE 350 $\mu$m data as a red layer. The main clumps and sub-regions of the field are marked. The green dashed polygons highlight the regions defined as L1688 and L1689+L1709 in the rest of the paper.}
\label{oph_rgb}
\end{figure*}

The Ophiuchus molecular cloud complex is a nearby region of low-mass star formation in the Gould Belt, located at a distance of $\sim$139${\pm 6}$ pc from the Sun (\citealp{2008AN....329...10M}). 
Using VLBA radio observations, \cite{2017ApJ...834..141O} derived a distance of $137.3 \pm1.2\,$pc for L1688 and $147.3 \pm 3.4\, $pc for L1689. 
\citet{2018ApJ...869L..33O} recently confirmed these distances with {\it Gaia} DR2 data, finding $138.4 \pm2.6\,$pc for L1688 and $144.2 \pm 1.3\, $pc for L1689. These recent VLBA and {\it Gaia} results  are in excellent agreement with our adopted distance of 139~pc. 
The stellar content of the Ophiuchus star-forming complex has been deeply studied for a long time at almost every wavelength of the spectrum,  revealing a rich environment of young stellar objects (YSOs), protostellar sources, and prestellar cores at various evolutionary stages \citep[e.g.,][and \citealp{Wilking+2008} for a review]{1989ApJ...340..823W, Leous+1991, Casanova+1995, 1998A&A...336..150M, 2015MNRAS.450.1094P}. 
The complex includes two concentrated clouds, L1688 and L1689,  extended by large-scale streamers stretching toward the northeast direction (\citealp{1990ApJ...365..269L}, see also Fig.~\ref{oph_rgb}). 
These streamers, visible with as dark lanes in optical photographs and detected in CO observations, have been part of the cobweb description of the Ophiuchus molecular cloud complex (\citealp{1989ApJ...338..902L, 1989ApJ...338..925L}). 
Despite being larger scale (e.g., more than 10 pc in length) than most filaments seen with {\it Herschel} in nearby clouds and being located outside the densest areas, L1688 and L1689, they provided the first direct evidence of the presence of filamentary structures in the complex.

While L1688 harbors dense star formation activity, L1689 is more quiescent (``the dog that didn't bark'', \citealp{2006MNRAS.368.1833N}). 
Despite its significant mass, L1689 does not have as much star formation activity as L1688 or other Gould Belt regions such as Aquila (e.g., \citealp{2010A&A...518L.102A}, \citealp{2015A&A...584A..91K}).
This peculiar distribution of star formation efficiency for independent, yet neighboring regions, sharing similar physical initial conditions, provides an interesting testbed to investigate the influence of potential triggers of star formation and the origin of stellar masses (cf. \citealp{1998A&A...336..150M}).

The molecular gas of the Ophiuchus star-forming complex is also well documented and has been scrutinized by ground-based and space-borne instruments for many years. 
In the present study, we use the naming conventions of \cite{1962ApJS....7....1L} and \cite{1990ApJ...365..269L} to refer to the different parts and clumps within the complex (cf. Fig.~\ref{oph_rgb}).
In particular, the clumpy structure of L1688 in the form of several dense gas clumps (Oph~A to Oph~F) was initially described by~\cite{1990ApJ...365..269L} based on DCO$^+$ observations. 
Oph~A is a very high extinction clump with several prominent prestellar and protostellar cores (\citealp{1993ApJ...406..122A}, \citealp{1998A&A...336..150M}), while Oph~B is a more quiescent region regarding dense core properties. 
Oph~C, E, F have lower column densities, but many infrared YSOs have been detected in this area, while Oph~D is an isolated quiescent clump with only a few objects (e.g., \citealp{1989ApJ...340..823W}, \citealp{2001A&A...372..173B}).

The whole Ophiuchus complex is under the heavy influence of, and feedback from, the Sco~OB2 association \citep[e.g.,][]{1986ApJ...306..142L,1998A&A...336..150M} enhancing the star formation activity in the L1688 cloud (\citealp{2006MNRAS.368.1833N}). 
While the young B3 star S1 may regulate the star formation activity locally in Oph~A by driving a compact HII region and an associated photodissociation region, the B2V star HD~147889 influences the cloud to the west of Oph A (\citealp{1973ApJ...184L..53G}, \citealp{1988ApJ...335..940A}, \citealp{1996A&A...315L.329A}). 
Furthermore, \cite{2006MNRAS.368.1833N} emphasized the influence on the cloud of the nearby O9V star $\sigma$ Scorpii (or $\sigma$ Sco for short) of the Sco OB2 association.

\begin{figure*}[h!]
\centering
\includegraphics[width=\hsize]{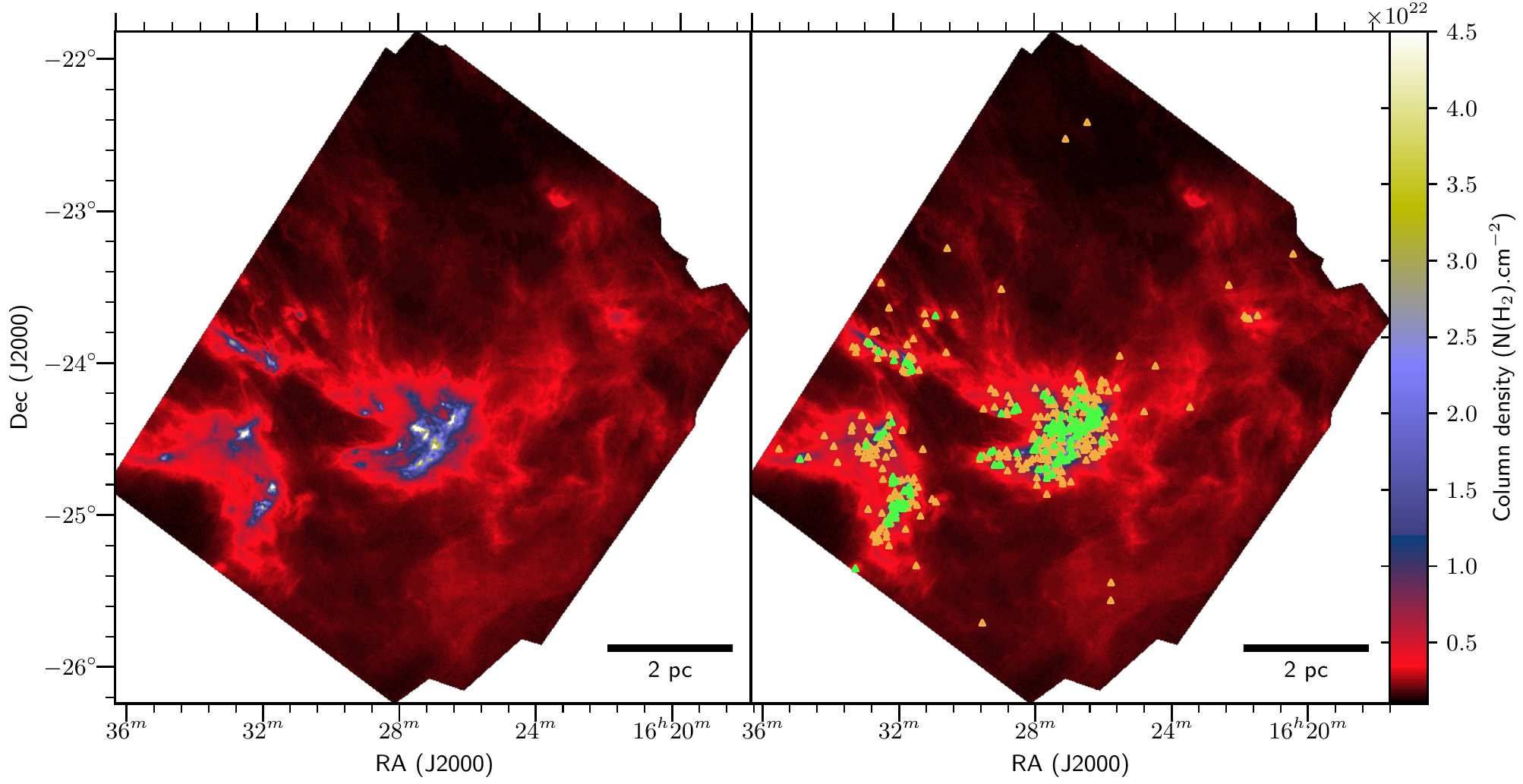}
\caption{{\bf (Left)} {\it Herschel} column-density map of the Ophiuchus molecular cloud at an effective HPBW resolution of  $18.2\arcsec $. 
The column-density map is given in units of H$_2$.cm$^{-2}$, calculated on a pixel-by-pixel basis usin a multi-scale graybody fitting procedure 
(see Appendix A of \citealp{2013A&A...550A..38P}).
{\bf (Right)} Same as the left panel with the candidate prestellar cores identified in Sect.~\ref{sec:corext} 
overlaid as green triangles and the unbound starless cores as orange triangles. 
The bar at the lower right indicates a scale of 2 pc.}
\label{fig:boundunboundfullmap}
\end{figure*}
\begin{figure}[!h]
\centering
\includegraphics[trim={0cm 0.5cm 0cm 1cm},clip,width=\hsize]{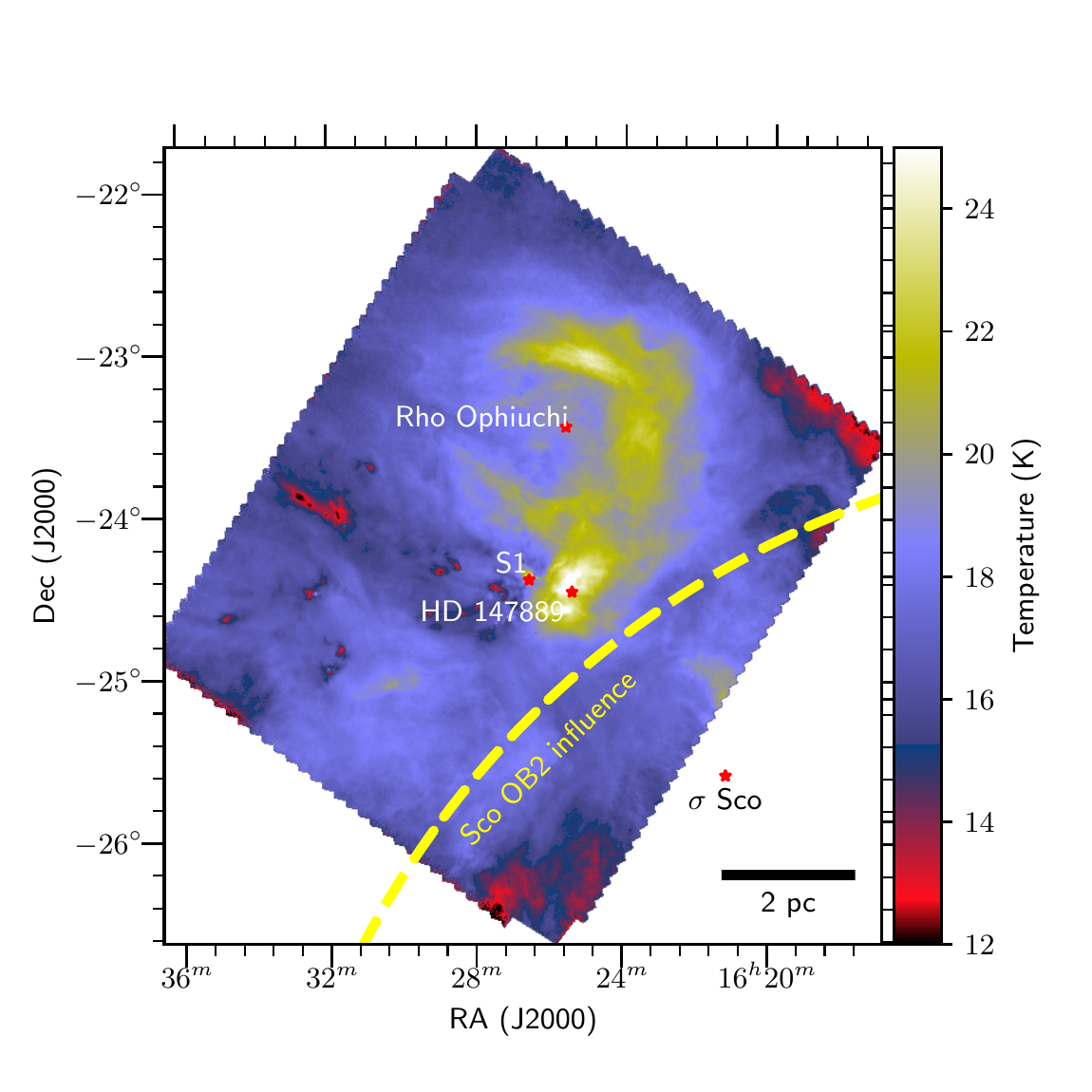}
\caption{{\it Herschel} dust temperature map of the Ophiuchus molecular cloud at a HPBW resolution of  $36.3\arcsec $ 
The dust temperature was derived by SED fitting from 160 $\mu$m to 500 $\mu$m. 
The warmest regions are located around S1, east of Oph~A, and HD~147889,  
while the warm shell to the northwest  is driven 
by the star  $\rho$~Ophiuchi. These three objects are B-type stars, S1 and HD~147889 being 
both young members of the L1688 infrared star cluster.}
\label{fig:ophtemperature}
\end{figure}

\section{Herschel observations and data reduction}
The {\it Herschel} Space Observatory (\citealp{2010A&A...518L...1P}) was the facility used in this study, performing broadband parallel-mode observations thanks to its two imaging cameras, PACS \citep{Poglitsch+2010} and SPIRE \citep{Griffin+2010}. 
The {\it Herschel} Gould Belt survey (HGBS)\footnote{http://gouldbelt-herschel.cea.fr} observations of L1688 and L1689 were taken on 25 sept. 2010, with the  {\it Herschel} Science Archive Observations IDs (OBSIDs): 1342205093, 1342205094. 
The SPIRE $250\, \mu$m data taken in parallel mode are saturated in the close vicinity of IRAS16293--2422 and a correction patch was observed in SPIRE-only mode, as OBSID 1342239773.
The SPIRE images used in the present study result from the combination of the parallel-mode data with the SPIRE-only patch around IRAS 16293-242 (Fig.~\ref{oph_rgb}).
The scanning speed of the parallel-mode observations was 60$\arcsec$.s$^{-1}$, with the same overall observing strategy as for every other HGBS region (\citealp{2010A&A...518L.102A}, \citealp{2015A&A...584A..91K}). 
For further reference and more details, the full processing pattern is described in \cite{2015A&A...584A..91K}.

\begin{table}[h!]
\caption{Observed {\it Herschel} bands and zero-level offsets}          
\label{table:1}    
\centering                     
\begin{tabular}{c c c c}       
\hline\hline              
Instrument &  Wavelength  & HPBW &  Map Offset \\    
 &  (microns)  & $\arcsec$ &  MJy/sr \\   
\hline                    
PACS & 70 & 8.4 & 4.7 \\    
PACS & 160 & 13.5 & 86.8 \\
SPIRE & 250 & 18.2 & 119.9 \\
SPIRE & 350 & 24.9 & 59.8 \\
SPIRE & 500 & 36.3 & 24.2 \\
\hline                                  
\end{tabular}
\end{table}
\subsection{PACS data reduction}
The parallel-mode PACS data at 70 $\mu$m  and 160 $\mu$m were reduced with HIPE (\citealp{2011ASPC..442..347O}), version 10, provided by the {\it Herschel} Science Center.
The path from the Level-0 data to the level-1 stage followed the standard steps of the {\it Herschel} parallel-mode pipeline, and flux calibration for PACS was applied using the last version of the correction factors (PACS CAL 45 0). 
A deglitching method was used  in HIPE to remove cosmic-ray hits on affected pixels.\@
High-level processing of the PACS data, including map making, was performed with \textsl{Scanamorphos} 20 (\citealp{2013PASP..125.1126R}) following~\cite{2015A&A...584A..91K} for scanning artifacts.
While the calibration error is estimated to be $\sim$5\% in our maps for point sources by default, we adopt a calibration error value of 10\% for the 70 $\mu$m band, and 20\% for the 160 $\mu$m band. The final products are maps reprojected on the same 3$\arcsec$-pixel grid at all wavelengths. 
The half-power beam width (HPBW) resolution of the maps  is given in Table~\ref{table:1} at each wavelength.

\subsection{SPIRE data reduction}
The reduction of the SPIRE maps was performed with HIPE version 10, where the scans in both orthogonal directions were combined. 
From level-0 to level-1, the standard calibration tree SPIRE CAL 10 1 provided by HIPE was used. We used the destriping module in HIPE for map correction by baseline subtraction. 
The initial maps had pixel sizes of 6$\arcsec$, 10$\arcsec$, and 14$\arcsec$ for the SPIRE 250 $\mu$m, 350 $\mu$m, and 500 $\mu$m bands, respectively. 
The maps were finally reprojected to a 3$\arcsec$-pixel grid for consistency with the PACS maps. 
The calibration error is estimated to be around 10\% for all SPIRE bands. 
\subsection{Saturation correction around IRAS 16293--2422}
With a flux of more than 50000 MJy/sr at 250 $\mu$m, IRAS 16293--2422, a Class~0 object in L1689 \citep[e.g.,][]{Walker+1986,Mundy+1986}, saturated the bolometers of SPIRE in the parallel-mode observations at 250 $\mu$m. 
A small field around this source was thus re-observed with {\it Herschel} using a different observing configuration for the SPIRE camera (OBSID: 1342239773). 
An image combination between the whole parallel-mode map of Ophiuchus and this particular patch was performed by detecting the saturation, which are pixels fixed at a $-99$ value, and replacing the saturated data by the data from the correction patch.

\begin{figure*}
\includegraphics[trim={0cm 0cm 0cm 1cm},clip,width=\hsize]{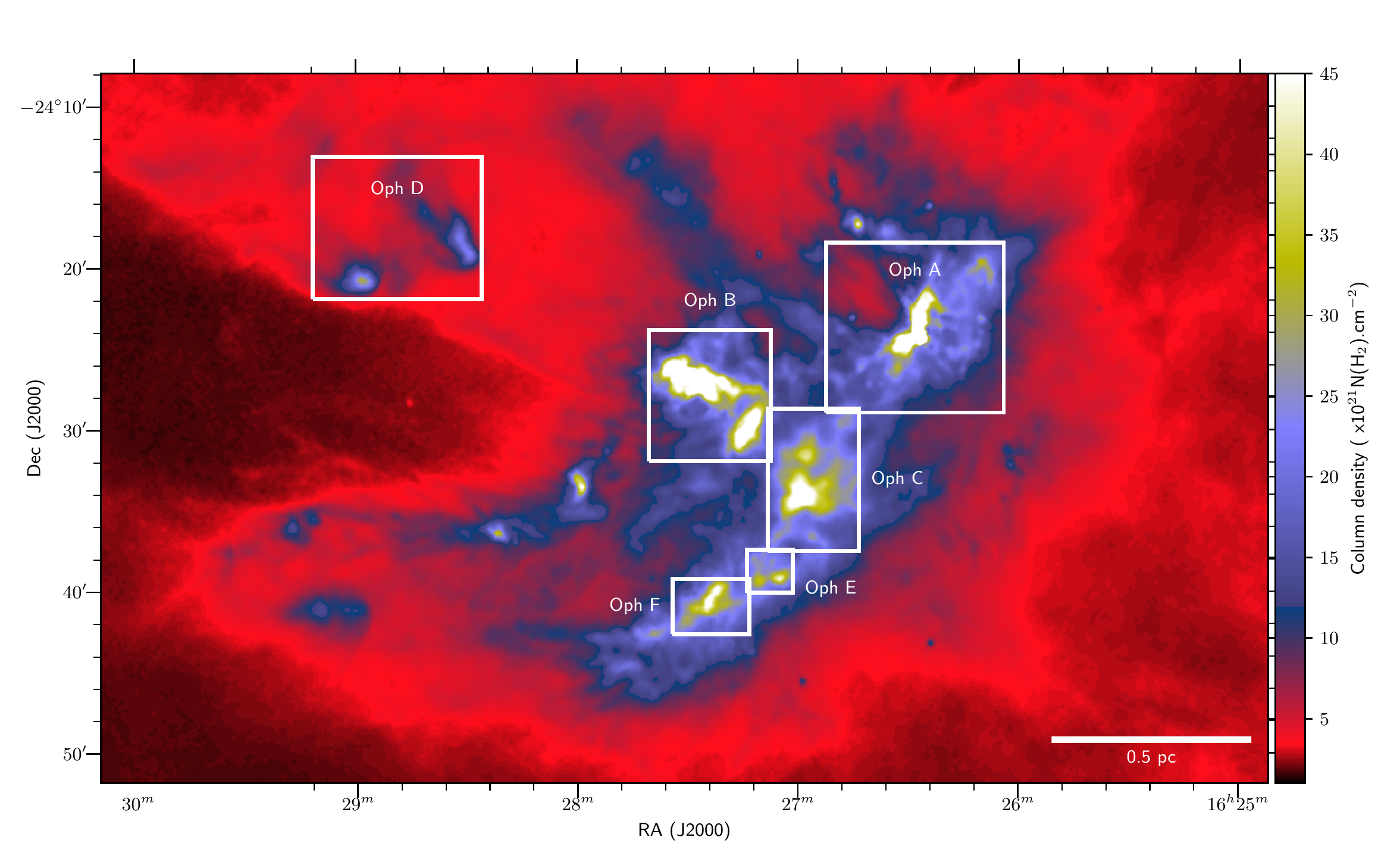}
\caption{Zoom on the {\it Herschel} column-density map of  Fig.~\ref{fig:boundunboundfullmap}a, 
showing the various dense clumps of L1688 named as in \cite{1990ApJ...365..269L}. 
The effective HPBW resolution is $18.2\arcsec $. The colorscale has been adjusted to transition from red to blue in the range of A$_V$ from 6 to 10.}
\label{oph_regions_l1688}
\end{figure*}
\begin{figure}
\includegraphics[trim={0cm 0cm 0cm 1cm},clip,width=\hsize]{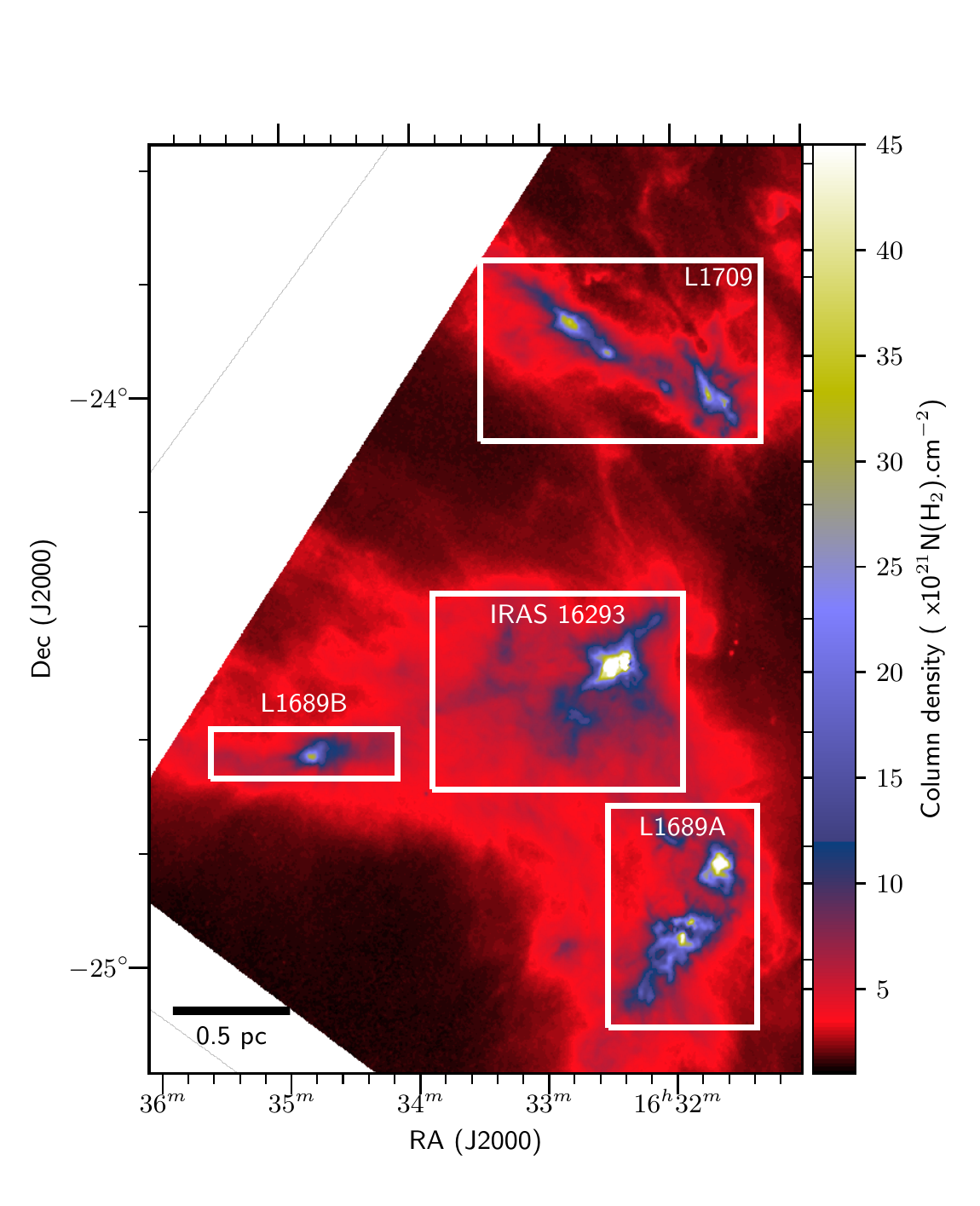}
\caption{Zoom on the {\it Herschel} column-density map of  Fig.~\ref{fig:boundunboundfullmap}, 
showing several sub-regions around L1689. The effective HPBW resolution is $18.2\arcsec $. The colorscale has been adjusted to transition from red to blue in the range of A$_V$ from 6 to 10.}
\label{oph_regions_l1689}
\end{figure}

\section{Results and analysis}
\subsection{Dust temperature and column density maps}
\label{sec:coldens}
The multi-wavelength {\it Herschel} data were used to derive a column density map and a dust temperature map for the Ophiuchus cloud. 
All  {\it Herschel} maps were first smoothed to a common $36.3\arcsec $ beam corresponding to the resolution of the SPIRE $500\, \mu$m data.  
Zero-level offsets, derived from {\it Planck} and IRAS data  (cf.~\citealp{2010A&A...518L..88B}), were added to the {\it Herschel} maps (see Table ~\ref{table:1} for the offset values). 
To obtain a column density map at a resolution of $36.3\arcsec $, we then fit a modified blackbody function to the 160 $\mu$m, 250 $\mu$m, 350 $\mu$m, and 500 $\mu$m data points on a pixel by pixel basis:
\begin{equation}
I_{\rm \nu} (x,y) = B_\nu[T_{\rm d} (x,y) ] \times \kappa_\nu \times \Sigma (x,y) ~,
\label{eq:blackbody}
\end{equation}
where I$_{\rm \nu}(x,y)$ is the surface brightness at frequency $\nu$ for pixel $(x,y)$, $T_{\rm d}(x,y)$ the dust temperature, $B_\nu[T_{\rm d}(x,y)]$ is the corresponding Planck function, $\kappa_{\rm \nu}$ the dust opacity per unit mass of dust$+$gas, and $\Sigma (x,y)$ the local gas$+$dust surface density of the cloud.

We also used a multi-scale decomposition scheme to add small-scale information from the SPIRE 350 $\mu$m/250 $\mu$m and PACS 160 $\mu$m data to the $36.3\arcsec$-resolution column-density map (see Appendix~A of \citealp{2013A&A...550A..38P}), obtaining  a ``high-resolution'' column-density map with an effective HPBW resolution of $\sim$18.2$\arcsec$ as a final result (see Fig.~\ref{fig:boundunboundfullmap}). 
Each point of the local spectral energy distribution (SED) was weighted by the absolute calibration error (20\% for PACS 160 $\mu$m  and 10\% for the SPIRE bands), and the same dust-opacity law as in other HGBS papers was assumed, namely $\kappa_{\lambda} = 0.1 \times (\lambda/\rm 300\mu m)^{-\beta}$ cm$^2$/g  as a function of wavelength in the submillimeter regime, with a dust emissivity index $\beta$ fixed to 2 (cf.~\citealp{1983QJRAS..24..267H}).
The same multi-scale combination scheme of \citet{2013A&A...550A..38P} was also adapted to merge the {\it Herschel} column-density data with the {\it Planck} optical depth map converted to column density using the HGBS dust opacity assumptions, as presented in Fig.~\ref{oph_rgb} (Bracco et al. in prep).

When smoothed to a common $5\arcmin $ resolution,  both the standard and the high-resolution column density map derived from {\it Herschel} data are consistent with a column density map derived from {\it Planck} data assuming the above-mentioned dust opacity law. 
More precisely, the column density ratio {\it Herschel}/{\it Planck} has a median value of 1.11, a lower quartile of 1.07, and an upper quartile of 1.16, for the material above $A_{\rm V} = 3$.

A similar comparison between our {\it Herschel} column-density map(s) and the 2MASS/NICER near-IR extinction map of \citet{2009A&A...493..735L}, both regridded and convolved to the same resolution, yields  a median ratio ({\it Herschel}/Extinction) of 0.65 in the range $2 \leq A_{\rm V} \leq 10$, assuming the standard conversion  $N_{\rm H_2}$(cm$^{-2}$) = 0.94 $\times$ 10$^{21}$ A$_V$ (mag) derived by \citet{1978ApJ...224..132B} for the diffuse interstellar medium.
Since in dense molecular clouds (typically for A$_{\rm V} \ga 6$), the conversion between A$_V$ and column density changes and becomes N$_{H_2}$(cm$^{-2}$) = 0.69 $\times$ 10$^{21}$ A$_{\rm V}$ (mag) (cf. \citealp{2003ARA&A..41..241D} and \citealp{2009ApJS..181..321E}), the calibration of our  column-density maps actually agrees with that of the 2MASS/NICER extinction map within $\sim \,$10\%--30\%.
We conclude that the absolute scaling of the {\it Herschel} column-density maps used in this paper is likely accurate to better than a factor of 1.5 (see also \citealp{2014A&A...562A.138R}).

\subsection{Distribution of mass in the Ophiuchus cloud}
\label{sec:ophcloudmass}
The detailed distribution of mass in the map of Fig.~\ref{fig:boundunboundfullmap} is closely related to the star formation process and other physical processes happening in the Ophiuchus molecular cloud, including the large-scale effect of the Sco OB2 association, and the shorter-scale effects of the shells formed by the stars $\rho$~Ophiuchi, HD 147889, and S1 (see Fig.~\ref{fig:ophtemperature} and \citealp{1998A&A...336..150M}). 
Our column-density maps allow us to directly trace the distribution of mass in the cloud, assuming it is the only significant cloud along the line of sight. 
This is a safe assumption given the position of the Ophiuchus molecular cloud above the Galactic plane (b = +16.5$^\circ$). 
We therefore calculated the mass of the Ophiuchus cloud enclosed within a given contour as 
\begin{equation}
M_{\rm cl} = \delta A_{\rm pixel}\,\mu_{\rm H_2}\, m_{\rm H} \times \sum\limits_{\rm pixels} N_{\rm H_2}, 
\end{equation}
where $\delta A_{\rm pixel} = (s_{\rm pixel}\times d)^2$ is the physical area covered by one pixel, a square of $s_{\rm pixel} =3\arcsec $ on a side at $d = 139\,$pc, 
$\mu_{\rm H_2}$ = 2.8  is the mean molecular weight per hydrogen molecule, m$_H$ is the hydrogen atom mass, and the sum is taken over the pixels inside the contour. 

In the following, we will define as dense molecular gas, the gas and dust material that lies above a column-density of 7 $\times$ 10$^{21}$ cm$^{-2}$ (i.e., A$_{\rm V} > 7$), the typical column-density level marking the transition between a regime of very low core and star formation efficiency
and a regime of high core and star formation efficiency in nearby clouds \citep[][]{2010A&A...518L.102A, 2010ApJ...724..687L, 2015A&A...584A..91K}. 
We obtained dense gas masses of $\sim$415 \msun~for L1688 (over the area shown in Fig~\ref{oph_regions_l1688}) and $\sim$165 \msun~for the complex formed by L1689 and L1709 (over the field shown in Fig~\ref{oph_regions_l1689}), both above 7 $\times$ 10$^{21}$ cm$^{-2}$. 
If we consider all of the gas, regardless of column density, the total masses become $\sim$980 \msun\ for L1688 and $\sim$890 \msun\ in L1689+L1709. 
A first observation is that the dense gas masses differ in the two sub clouds by more than a factore of two, but the total gas masses are on par with each other. The material reservoir is therefore the same in both clouds, but they have different compression status. 

\begin{figure*}[!h]
\centering
\includegraphics[width=0.49\hsize]{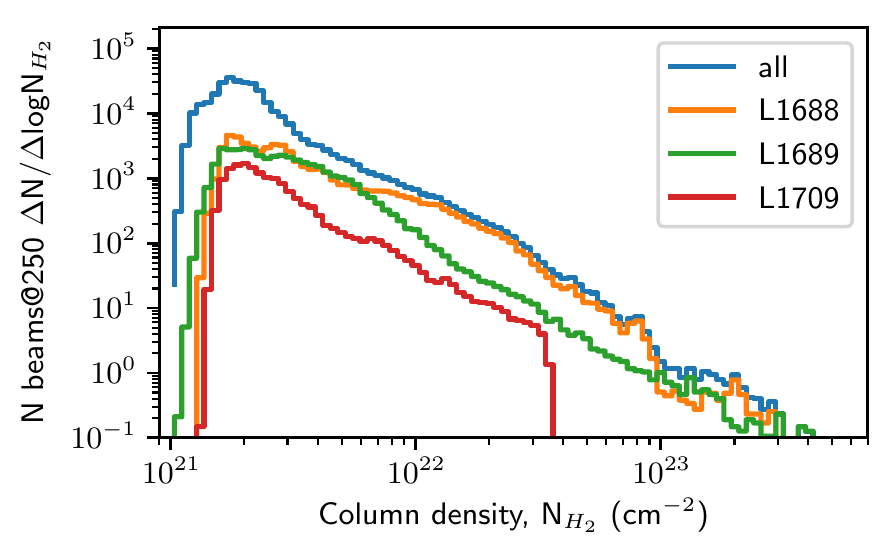}
\includegraphics[width=0.49\hsize]{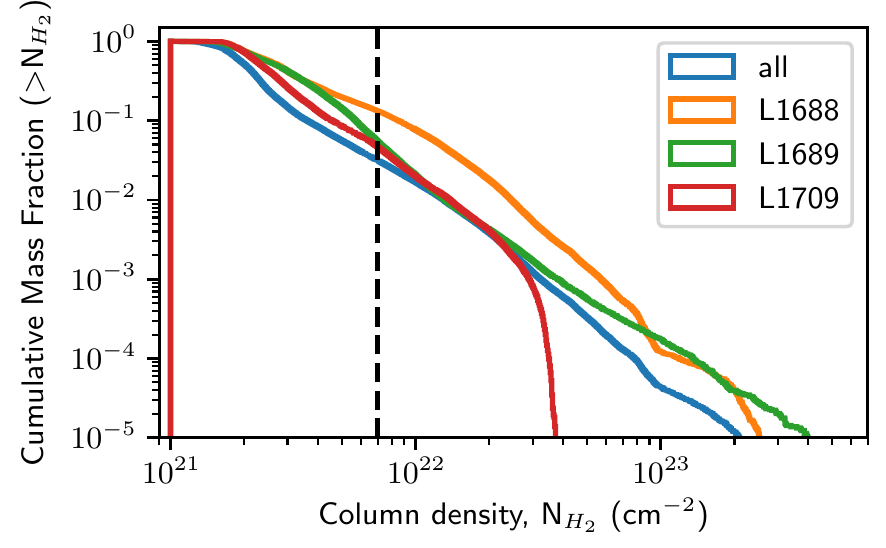}
\caption{{\bf (Left)} Probability density function (PDF) of column density in the Ophiuchus molecular cloud and its subclouds L1688, L1689 and L1709, as derived from the high-resolution column density map at a resolution of 18$\arcsec$2. (See Table~\ref{table:2} for the results of power-law fits to the high-column density tails of the PDF.)  
The PDF of L1688 shows an overdensity around $N_{\rm H_2} \sim 2\times 10^{22}\, {\rm cm}^{-2}$, probably as a consequence of the Sco~OB2 influence.
{\bf (Right)} Normalized cumulative mass fraction as function of column density for the Ophiuchus molecular cloud using the same column-density map as in the top figure. 
The dashed line represents the fiducial dividing line between dense star-forming gas and lower-density gas. 
In the various parts of the cloud, the vast majority of the gas is found at low column density (A$_{\rm V}$ $\leq$ 2).}
\label{fig:PDF}
\end{figure*}

Given the column density map, we can derive the probability density function (PDF) of column density in the cloud and display this function separately for each sub-cloud (Fig.~\ref{fig:PDF}). 
While the low column density part of the PDF is usually well fit by a lognormal distribution, either due to turbulence (\citealp{2009A&A...508L..35K}, \citealp{2013ApJ...766L..17S}), or to incompleteness effects (\citealp{2017A&A...606L...2A}), we find that the situation in Ophiuchus is more complex due to the presence of various star-forming cloud environments, each contributing separately to the PDF.\@ 
Table~\ref{table:2} provides the results of piecewise power-law fits to the PDF of each subcloud in Ophiuchus. 
The break points of each part of the PDF were defined by fitting two power-laws in the high column density tail above a fiducial level of A$_V = 7$, which again defines the dense gas within which most prestellar cores (\citealp{2010A&A...518L.102A}; \citealp{2015A&A...584A..91K}) and young stars (e.g., \citealp{2010ApJ...724..687L}; \citealp{2010ApJ...723.1019H}) are observed in nearby clouds.

A simple relationship exists between the high mass slope of the PDF derived in a region, and the associated density profile (see, e.g., \citealp{2013ApJ...763...51F}). 
For a spherical cloud with a density profile $ \rho \propto r^{-\alpha}$, that is, a column density profile $ \Sigma \propto r^{-\alpha+1}$, the logarithmic slope $s$ of the column density PDF, defined by $\frac{dN}{dlogN_{\rm H_2}} \propto N_{\rm H_2}^s$ can be expressed as:
\begin{equation}
  \alpha_{sph} = -\frac{2}{s}+1, 
\end{equation}
while for a cylindrical cloud the logarithmic slope is:
\begin{equation}
  \alpha_{cyl} = -\frac{1}{s}+1. 
\end{equation}
Given the morphology of the cloud, with multiple clumps gathered in a single, large molecular cloud, we know that the overall PDF is the sum of  contributions from individual clumps. 
It remains difficult to characterize the column-density PDF of each clump separately given the relatively low number of independent beams per clump.
The typical values reported in Table~\ref{table:2} for the power-law fits to the PDF are compatible with $\alpha_{sph}$ = 2 and $\alpha_{cyl}$ = 1.5, which is physically reasonable for a self-gravitating cloud and consistent with results found in other regions \citep[e.g.,][]{2012A&A...540L..11S, 2013ApJ...766L..17S, 2015MNRAS.453.2036B}.

\begin{table*}[!h]
\footnotesize
\caption{Results of power-law fits to the column density PDFs in the Ophiuchus subclouds. }  
\label{table:2}    
\centering                        
\begin{tabular}{c c c c c c c}     
\hline\hline             
Region      & Range                                                          & Power law fit$^\dag$            & Cumulative power law fit$^\dag$     &   $\alpha_{sph}$  &   $\alpha_{cyl}$    \\
            &                        (cm$^{-2}) $                                      & dN/dlog$N_{\rm H_2} \propto$  & $M(>N_{H_2}) \propto$                  &                   &                     \\
\hline                     
L1688 (single PL fit$^\dag$$^\dag$) & 7$\times$10$^{21} $ $\leq$ N(H$_2$)                                             & N$_{H_2}$$^{-2.5\pm0.07}$  & N$_{H_2}$$^{-2.7\pm0.04}$               &      1.8         &      1.5             \\\hdashline
L1688 (broken PL fit$^\dag$$^\dag$) & 7$\times$10$^{21} $ $\leq$ N(H$_2$) $\leq$ 1.7$\times$10$^{22}$                 & N$_{H_2}$$^{-1.3\pm0.10}$  & N$_{H_2}$$^{-1.8\pm0.05}$                &      2.5         &      1.8             \\
      & 1.7$\times$10$^{22}$ $\leq$ N(H$_2$) $\leq$ 7.8$\times$10$^{22}$                & N$_{H_2}$$^{-2.7\pm0.09}$  & N$_{H_2}$$^{-2.8\pm0.02}$               &      1.7         &      1.4             \\
\hline                    
L1689 (single PL fit) & 7$\times$10$^{21} $ $\leq$ N(H$_2$)                                             & N$_{H_2}$$^{-2.1\pm0.04}$  & N$_{H_2}$$^{-2.1\pm0.02}$               &      2.0         &      1.5             \\\hdashline
L1689 (broken PL fit) & 7$\times$10$^{21} $ $\leq$ N(H$_2$) $\leq$ 4.9$\times$10$^{22}$                 & N$_{H_2}$$^{-2.4\pm0.07}$  & N$_{H_2}$$^{-2.3\pm0.03}$               &      1.8         &      1.4             \\
      & 4.9$\times$10$^{22} $ $\leq$ N(H$_2$)                                           & N$_{H_2}$$^{-1.8\pm0.10}$  & N$_{H_2}$$^{-2.0\pm0.04}$                &      2.1         &      1.6             \\
\hline                      
L1709 (singe PL fit) & 7$\times$10$^{21} $ $\leq$ N(H$_2$)                                             & N$_{H_2}$$^{-2.1\pm0.06}$  & N$_{H_2}$$^{-2.7\pm0.10}$              &      2.0         &      1.5             \\
\hline                               
\end{tabular}
\tablefoot{\tablefoottext{\dag}{Power-law fits to both the differential and the  cumulative PDF were performed for each column-density range given in Col.~2. \newline
\tablefoottext{\dag\dag} The results of single power-law (PL) fits are given for the whole range of column densities 
above N$_{H_2}$ $\sim $ 7$\times$10$^{21}$ cm$^{-2}$.  Broken PL fits  are also provided 
for the L1688 and L1689 PDFs. The break points were left free and identified using 
the Multivariate Adaptive Regression Splines technique (\citealp{10.2307/2241837}) via its python implementation, py-earth.
}} 
\end{table*}

\subsection{Filamentary texture of the Ophiuchus molecular cloud}
\begin{figure}[!h]
\centering
\includegraphics[trim={0.5cm 0cm 0cm 1cm},clip,width=\hsize]{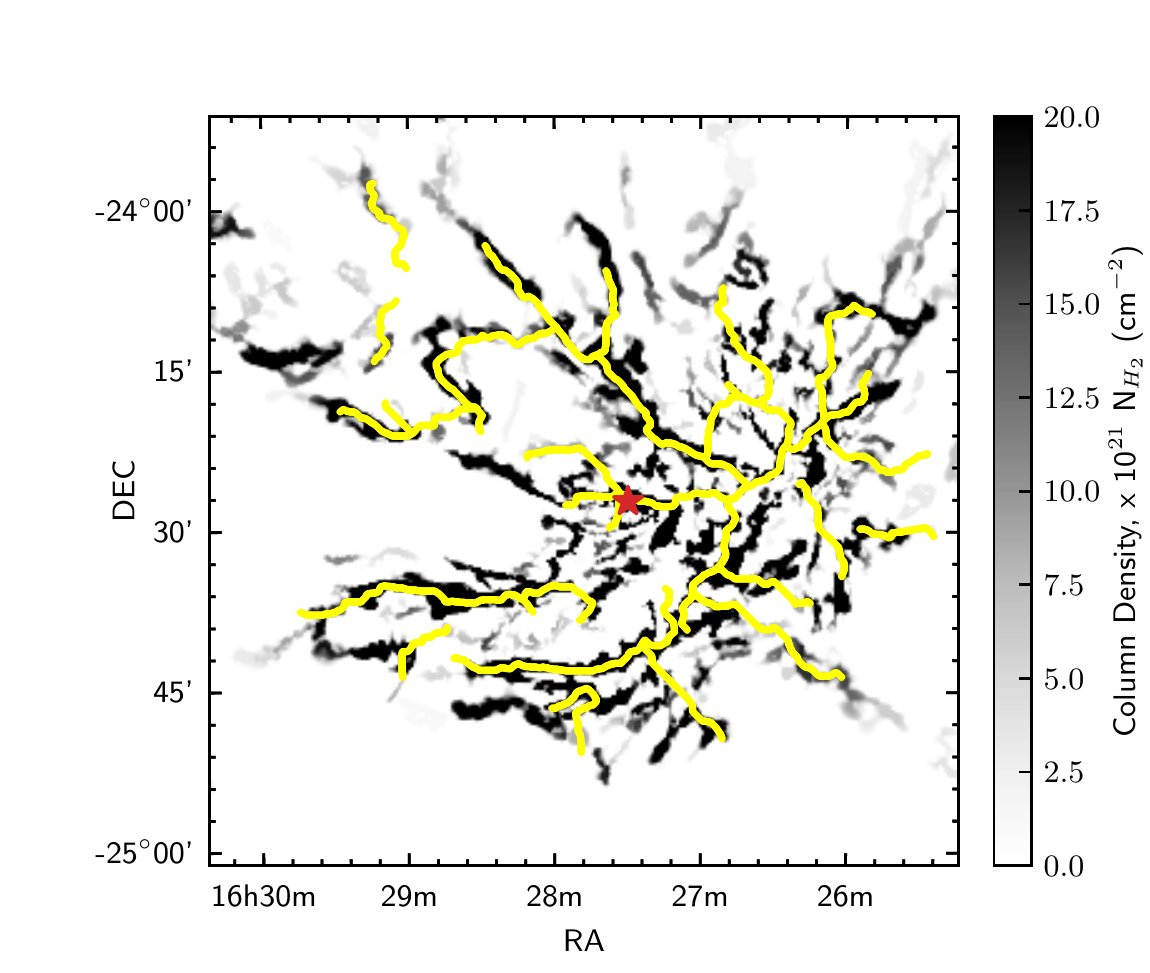}
\includegraphics[trim={0.5cm 0cm 0cm 1cm},clip,width=\hsize]{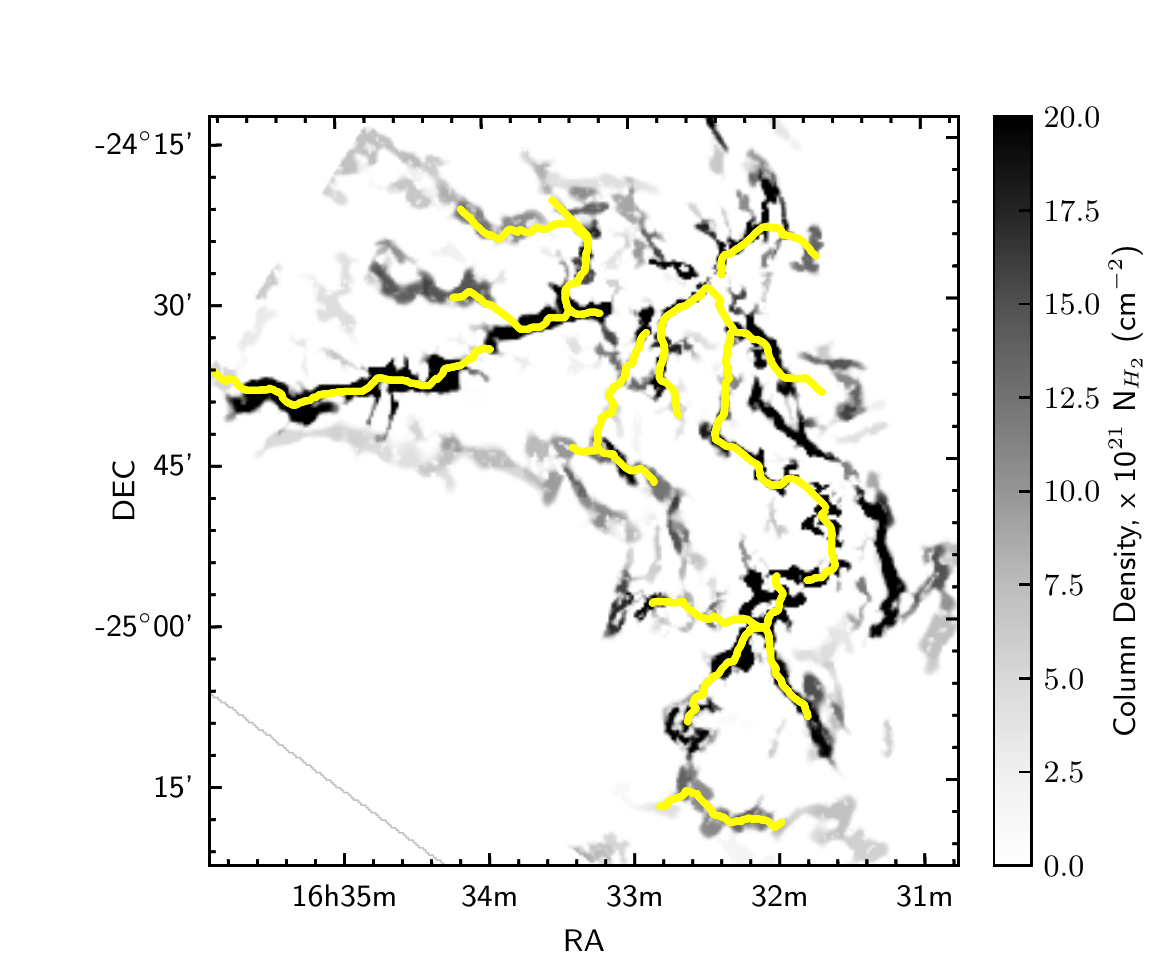}
\caption{Filtered versions of the \textit{Herschel} column density maps of 
L1688 {\bf (top)} and L1689 {\bf (bottom)}, where the contrast of filamentary structures has been enhanced 
by accumulating small-scale emission up to a transverse scale of $150\arcsec $ ($\sim$ 0.1 pc) using the \textsl{getfilaments} algorithm (see Sect.~4.3). 
The grayscale represents 
effective column density 
within the skeleton masks generated by \textsl{getfilaments}. 
The skeleton of the filaments independently extracted with the DisPerSE algorithm is displayed in yellow.
In the top panel, the red star marks the center of mass of L1688.}
\label{fig:boundunboundfullfilmap}
\end{figure}

The Ophiuchus cloud reveals a wealth of extended filamentary structure at  parsec scale when observed with the high sensitivity and dynamic range of \textit{Herschel} (Fig.~\ref{oph_regions_l1688}, Fig.~\ref{oph_regions_l1689}). 
Using \textsl{getfilaments} (\citealp{2013A&A...560A..63M}), we extracted and identified the main filamentary structures in the mapped area. 
Based on a multi-scale filtering approach, \textsl{getfilaments} can separate filaments from the background and measure their properties.
Filtered images accumulating transverse angular scales up to an upper limit of interest can be constructed. 
In Fig.~\ref{fig:boundunboundfullfilmap}, we adopted an upper angular scale of 150$\arcsec$, corresponding to a linear scale of $\sim 0.1\,$pc, which is the typical inner width of {\it Herschel} filaments in nearby clouds \citep[][]{Arzoumanian+2011, Arzoumanian+2019}. 
We independently extracted filaments in the Ophiuchus region using the DisPerSE algorithm (\citealp{2011MNRAS.414..350S}) with a persistence threshold of 0.2 $\times$ 10$^{21}$ cm$^{-2}$, a robustness threshold of 2 $\times$ 10$^{21}$ cm$^{-2}$, and an assembling angle of 65 $^{\circ}$. 
DisPerSE allows the filaments to be reconnected and filtered by "persistence", while \textsl{getfilaments} enhances the contrast of all filamentary structures present  in the cloud via a multi-scale decomposition. 
The observation of widespread filamentary structures in L1688 and L1689 is quite remarkable since these two centrally-condensed clouds are not known to be filamentary in the literature, in contrast to the elongated streamers, L1709 and L1712 (cf. \citealp{1989ApJ...338..902L, 1989ApJ...338..925L}).

\subsection{Core extraction with \textsl{getsources}}
\label{sec:corext}
\subsubsection{Multiwavelength detections and measurements with \textsl{getsources}}

To identify compact emission sources corresponding to candidate dense cores within the structured background emission traced by {\it Herschel} at all wavelengths, we used the source extraction algorithm \textsl{getsources}\footnote{The HGBS first-generation catalog of cores presented in this paper (see Appendix~\ref{sec:appendix_catalog}) was produced with the ``November 2013'' major release of \textsl{getsources} (v1.140127), which is publicly available from http://gouldbelt-herschel.cea.fr/getsources.} (\citealp{2012A&A...542A..81M}). 
This multi-wavelength, multi-scale algorithm has been adopted for all HGBS ``first-generation'' catalog studies (e.g., \citealp{2015A&A...584A..91K},~\citealp{2016MNRAS.459..342M}, \citealp{2018A&A...615A.125B}). The algorithm extracts sources from the data in two main steps, first a detection step and second a measurement step.
These detection and measurement steps are fully described in~\cite{2012A&A...542A..81M} and~\cite{2015A&A...584A..91K}, and the processing has been standardized for easier use in every HGBS study. 

For source extraction, we used the {\it Herschel} maps in the five bands at 70, 160, 250, 350, and 500 $\mu$m, as well as the high resolution column-density map ($18.2\arcsec$ resolution) and a temperature-corrected 160~$\mu$m map. 
The latter map was obtained by converting the original 160 $\mu$m map to column density assuming the dust temperature given by the color-temperature ratio between 160 $\mu$m and 250 $\mu$m.
We performed two sets of source extractions, one based on detections between 160 $\mu$m and 500 $\mu$m to identify  dense cores, and the other based on detections at 70 $\mu$m to identify YSOs including protostellar objects.
Although the extraction process is fully automated and the  \textsl{getsources} code itself performs a pre-selection of sources, several additional criteria remain crucial for selecting reliable core detections (see next subsection). 

\subsubsection{Automatic selection and derivation of physical properties}

Following the \textsl{getsources} extractions, we applied additional selection criteria to ensure the reliability and physical robustness of the catalog of candidate dense cores returned by \textsl{getsources} \citep[cf.][]{2015A&A...584A..91K}:

\begin{itemize}
\item The global detection significance over all bands [see Eq~(18) of \citealp{2012A&A...542A..81M}] 
must be greater than 10;
\item The column density detection significance must be greater than 5 in the high-resolution column density map;
\item The global goodness parameter must be greater than 1;
\item The measurement signal-to-noise ratio must be greater than 1 in the high-resolution column-density map;
\item The monochromatic detection significance must be greater than 5 in at least two bands between 160 $\mu$m and 500 $\mu$m, 
and the measurement signal-to-noise ratio must be greater than 1 in at least one of these bands.\\ 
We refer the reader to \citet{2012A&A...542A..81M} for the exact definition of "significance" and "goodness" for \textsl{getsources}.
\end{itemize}

As a final, post-selection check, the 740 candidate cores automatically selected on the basis of these criteria were all inspected visually.
The main properties of each dense core in the resulting,  final catalog were then derived by SED fitting (see Sect.~4.5 below). 
After a first pass to fit a simple modified blackbody model to the photometric data points provided by the \textsl{getsources} extractions, we selected a total of 513 candidate dense cores, including 459 starless cores and 54 protostellar cores or more evolved objects (e.g., Class~II YSOs).
Protostellar cores and YSOs were identified based on the detection of point-like emission at 70 $\mu$m \citep[cf.][]{Dunham+2008}. 
In the rest of this paper, we are not discussing evolved YSOs since they are not closely associated with dense cores.

\subsection{Dense core properties}

\begin{figure}[!htp]
\centering
\includegraphics[width=0.90\hsize]{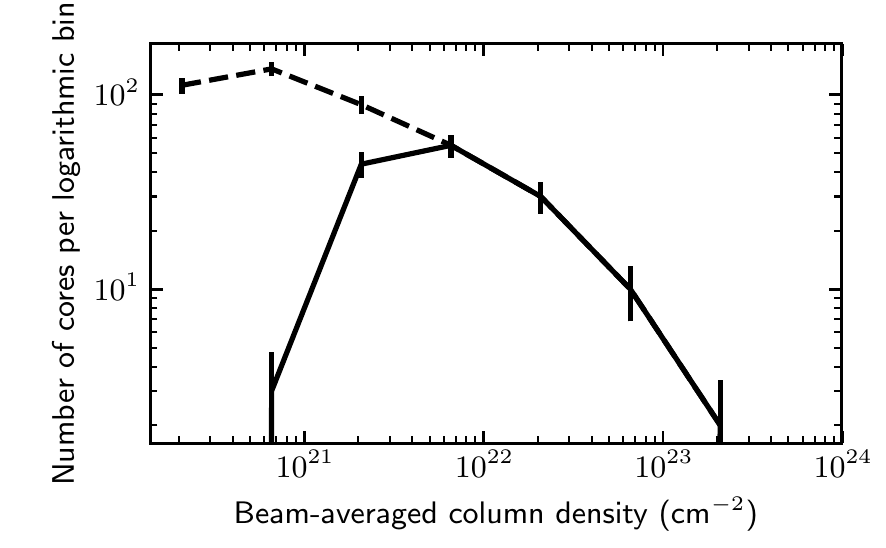}
\includegraphics[width=0.90\hsize]{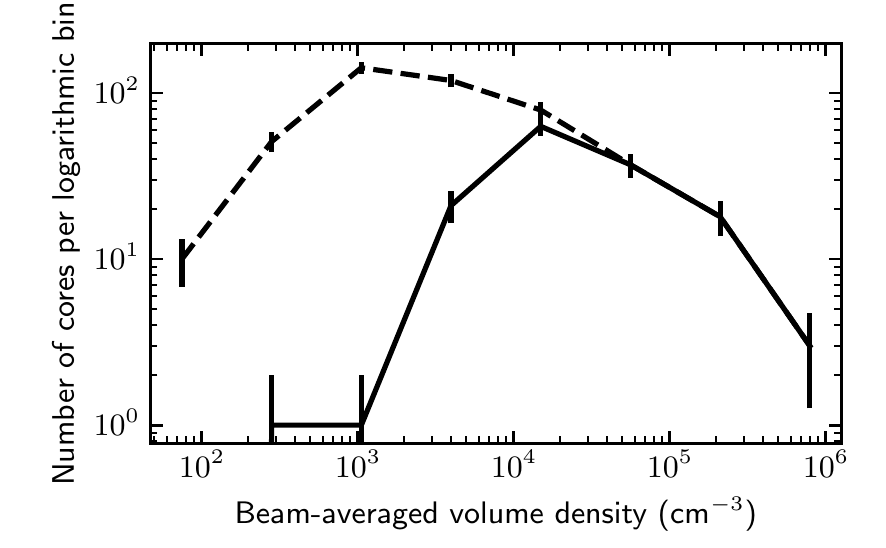}
\caption{Distributions of beam-averaged column densities {\bf (top)} and beam-averaged volume densities {\bf (bottom)}, at the resolution of the 
SPIRE 500\,$\mu$m data,
for the populations of 459 starless cores (dashed) and 144 candidate prestellar cores (solid) identified in the Ophiuchus complex.}
\label{fig:colvol_dens}
\end{figure}

Exploiting the wavelength coverage of \textit{Herschel} and multi-wavelength capability of \textsl{getsources}, we constructed the SED of each source from the integrated flux densities measured at 160--500\,$\mu$m. 
By fitting these SEDs with a modified blackbody function, in a manner similar to the method employed for deriving the column-density map (cf. {Sect.~\ref{sec:coldens}} and Eq.~\ref{eq:blackbody}), we obtained estimates of the average line-of-sight temperature and mass of each candidate core. 
For consistency with other HGBS studies, we derived a deconvolved radius for each source, calculated as the square root of the quadratic difference between the geometric mean of the major/minor angular sizes and the HPBW size of the high-resolution column density map ($18.2\arcsec$). 
For each core, we also computed the geometric mean of the major and minor FWHM sizes, the peak column density, the beam-averaged column density, the central beam volume density, and the average volume density (cf. Fig.~\ref{fig:colvol_dens} for their distributions), assuming a Gaussian density distribution within each object \citep[cf.][]{2015A&A...584A..91K}. 
All of these quantities are provided in online Table~\ref{tab_der_cat_cores} (see Appendix~A for a sample portion). 

From the derived properties, we selected self-gravitating prestellar cores among the group of starless cores based on a criterion related to the critical Bonnor-Ebert (BE) mass. 
The latter can be expressed as:

\begin{equation}
  M_{BE, crit}  \approx  \frac{2.4\times R_{BE}\, c_s^2}{G},
\end{equation}

\noindent where R$_{BE}$ is the BE radius, c$_s \approx 0.19\,$km/s the isothermal sound speed (assuming a temperature of $\sim 10\,$K for the dense gas in the ambient cloud), and G the gravitational constant (\citealp{1956MNRAS.116..351B}). 
We estimated the outer radius, $R_{\rm obs}$, of each core from the FWHM source diameter measured in the high-resolution column-density map, deconvolved from the effective HPBW resolution of the map (18.2$\arcsec$).
We then compared the mass derived from {\it Herschel} observations, $M_{\rm obs}$, to the critical BE mass estimated as $M_{BE, crit} = 2.4\times R_{\rm obs}\, c_s^2/G $, and selected self-gravitating cores based on the evaluated BE mass ratio: 
\begin{equation}
  \alpha_{BE}  \equiv  \frac{M_{BE, crit}}{M_{\rm obs}} \leq 2.
\end{equation}

\begin{figure*}[!htp]
\centering
\includegraphics[width=\hsize]{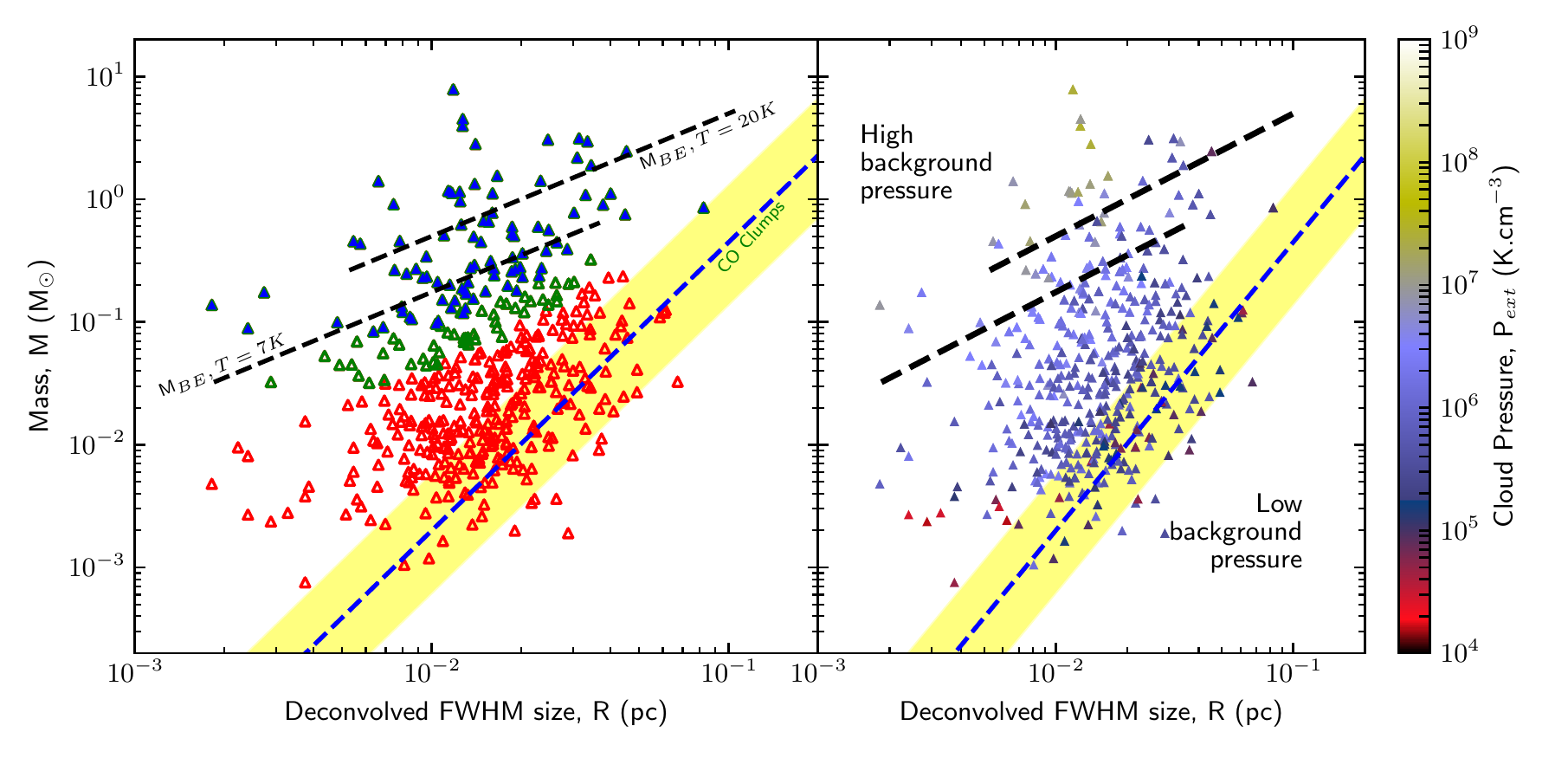}
\caption{ {\bf (Left) } Mass versus size diagram for the entire population of 459 starless cores identified with {\it Herschel} in the Ophiuchus molecular cloud. 
The core FWHM sizes were measured with \textsl{getsources} in the high-resolution column density map (Fig.~\ref{fig:boundunboundfullmap}) and deconvolved from an 18$\arcsec$2 (HPBW) Gaussian beam. 
The core masses were derived via graybody fitting to the \textsl{getsources}-estimated fluxes (see text). 
The 93 robust prestellar cores (for which $\alpha_{BE} \le 2 $) are shown as filled blue symbols, the other (candidate) prestellar cores as open green symbols, and the other (unbound) starless cores as open red symbols. For comparison, models of critical isothermal Bonnor-Ebert spheres at T = 7 K and T = 20 K are plotted as dashed black lines. 
The mass-size correlation observed for diffuse CO clumps \citep[cf.][]{1996ApJ...471..816E} is displayed as a shaded yellow band. 
{\bf (Right)} Same mass versus size diagram for all 459 starless cores, where the color coding of the data points is related to the ambient cloud pressure, $P_{{\rm cl}} \approx 0.88~G~\Sigma_{\rm cl}^2$, estimated from the local background column density $\Sigma_{\rm cl}$ assuming a cloud in virial equilibrium \citep{2003ApJ...585..850M}.}
\label{fig:regboundunboundcoresmasssize}
\end{figure*}

This provided a first set of prestellar cores, hereafter called robust prestellar cores, with a total of 93 objects after visual post-selection checks (see filled blue triangles in the top panel of  Fig.~\ref{fig:regboundunboundcoresmasssize}). 
We also used an alternative, less stringent criterion to select candidate prestellar cores among starless cores, corresponding to the empirical lower envelope of self-gravitating objects in the mass--size diagram of Fig.~\ref{fig:regboundunboundcoresmasssize} according to Monte-Carlo simulations similar to those described in Sect.~\ref{sec:completeness} below and Appendix~B \citep[see][]{2015A&A...584A..91K}:

\begin{equation}
  \alpha_{BE} \le 5 \times (HPBW_{N_{\rm H_2}}/FWHM_{N_{\rm H_2}})^{0.4},
\end{equation}

\noindent 
where HPBW$_{N_{\rm H_2}} = 18.2\arcsec$ is the HPBW resolution of the column density map and FWHM$_{N_{\rm H_2}}$ is the FWHM size of the source measured in the column density map. 
This second criterion allows us to consider as candidate prestellar cores unresolved objects whose physical radius is more uncertain than that of resolved cores and which do not satisfy Eq.~(6).
Using this alternative criterion, 51 additional candidate prestellar cores were selected (see open green triangles in the top panel of Fig.~\ref{fig:regboundunboundcoresmasssize})
To summarize, we identified a total of 144 candidate prestellar cores satisfying Eq.~(7), including 93 robust prestellar cores matching the strict Bonnor-Ebert criterion of Eq.~(6).
The observed and derived properties of all dense cores are given in two accompanying online catalogs (cf. Tables~A.1 and A.2 in Appendix~A, respectively).

The bottom panel of Fig.~\ref{fig:regboundunboundcoresmasssize} shows that there is an interesting correlation between the masses of the cores and the local pressure in the ambient cloud, $P_{{\rm cl}}$. 
Assuming the ambient cloud is in approximate virial equilibrium, the latter can be estimated as $P_{{\rm cl}} \approx 0.88~G~\Sigma_{\rm cl}^2$ \citep{2003ApJ...585..850M}, where $\Sigma_{\rm cl} \equiv \mu_{H_2}\, m_H\,  N_{\rm H_2}^{\rm cl} $ is the local background cloud column density of each core as estimated from {\it Herschel} data with \textsl{getsources}. 
As can be seen in Fig.~\ref{fig:regboundunboundcoresmasssize}, there is a clear trend for most self-gravitating prestellar cores to lie in high-pressure, high-column-density parts of the cloud and for low-mass unbound starless cores to lie in low-pressure areas. 
We will come back to this trend in Sect.~5 below in connection with the spatial distribution of the various subsets of dense cores.

\subsection{Completeness of the prestellar core survey}\label{sec:completeness}

Monte-Carlo simulations were performed to estimate the completeness of our census of prestellar cores in the Ophiuchus region (see Appendix~B). 
Clean background emission maps of the L1688 region were first constructed at all {\it Herschel} wavelengths (including a column density plane), by subtracting the emission of all compact sources identified with \textsl{getsources} from the observed maps.
A population of 285 model Bonnor-Ebert-like cores were then inserted throughout the clean-background images to generate a full set of synthetic {\it Herschel} and column density images for L1688. 
The model cores were given a flat mass distribution such as $\Delta N/\Delta$log$M$ $\propto$ $M^{0}$ from $0.05\, M_\odot$ to $2.0\, M_\odot$.
An $M \propto R$ mass versus size relation appropriate for isothermal Bonnor-Ebert  spheres was adopted. 
The dust continuum emission from the synthetic Bonnor-Ebert cores in all {\it Herschel} bands was simulated using an extensive grid of spherical dust radiative transfer models constructed by us with the MODUST code (e.g., \citealp{2000A&A...360..213B, 2001A&A...375..950B}).

Based on the results of these simulations, our {\it Herschel} census of candidate prestellar cores in Ophiuchus is estimated to be $> 80\% $ complete above a true core mass of $\sim$0.1~$M_\odot$ (cf. Fig.~\ref{fig:complete}), corresponding to an observed core mass of $\sim$0.2~$M_\odot$ on average given that observed masses appear to be typically underestimated by a factor of $\sim 2$ near the completeness limitdue to the internal temperature gradient within starless cores (see top and bottom panels of Fig.~\ref{fig:complete2} in Appendix~B).

The simple model of the core extraction process described in Appendix~B of \citet{2015A&A...584A..91K} was also used to independently assess the completeness level and its dependence on the local column density of the ambient cloud.
This model shows that the completeness of prestellar core extractions does decrease as background cloud column density and cirrus noise increase but suggests that the global completeness curve of the prestellar core survey in Ophiuchus is consistent with that inferred from the Monte-Carlo simulations (Fig.~\ref{fig:complete}).

\section{Discussion}

\subsection{A bimodal distribution of filament orientations}
\begin{figure}[!h]
\centering
\includegraphics[width=\hsize]{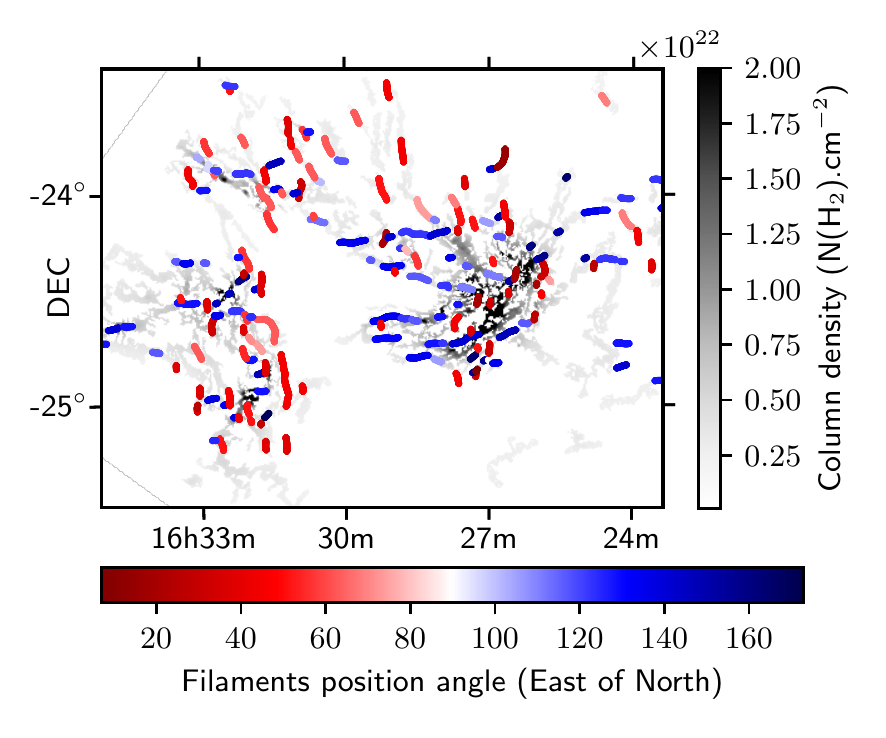}
\includegraphics[width=\hsize]{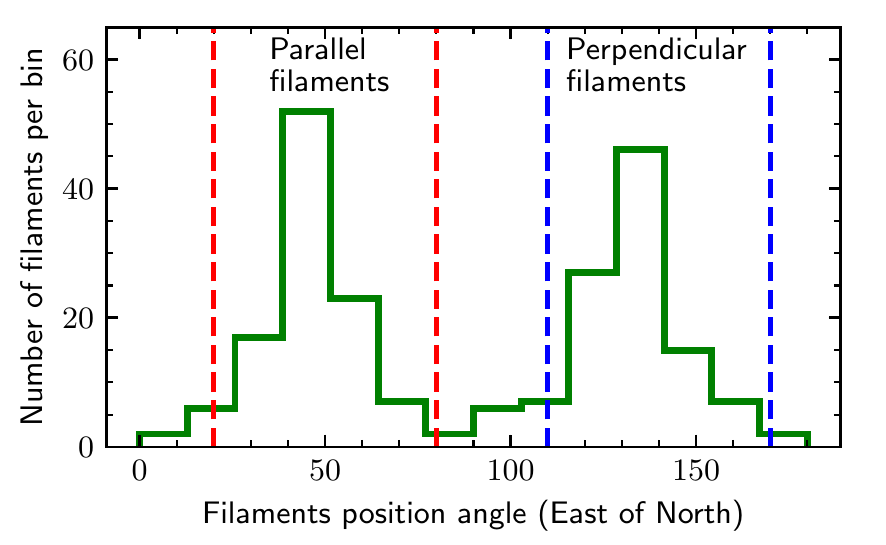}
\caption{{\bf (Top) } Skeleton map of filament orientations (colorscale) overlaid on a filtered version of the 
\textit{Herschel} column density map of the Ophiuchus cloud (grayscale) obtained with the  \textit{getfilaments} algorithm (see Fig.~\ref{fig:boundunboundfullfilmap}). 
The filamentary structures identified with \textsl{getfilaments} are displayed in red or blue depending on their position angles measured east of north (see colorscale).
{\bf (Bottom) } Distribution of median position angles (P.A.) for the sample of filaments identified with \textsl{getfilaments} 
(cf. Sect.~4.3). 
Two modes of filament orientations are clearly visible. 
The ``parallel''  filaments (with P.A. from 20 to 80 degrees east of north) are following the general direction of the large-scale streamers (cf. Fig.~\ref{oph_rgb}) and may be associated with material swept-up by the winds of the Sco~OB2 association (whose general direction is about 45 degrees east of north), while the ``perpendicular''  filaments (with P.A. from 100 to 160 degrees) may have formed as a result of large-scale compression from the Sco~OB2 winds.  
}
\label{fig:filamentorientation}
\end{figure}

The shape of the Ophiuchus molecular cloud seems to be organized as a hub of filaments converging toward the center of mass of the main cloud L1688 as marked by a red star in  Fig.~\ref{fig:boundunboundfullfilmap}. 
One of the interesting features of this region is the U-shape of the densest parts of the cloud (cf. Fig.~\ref{fig:subregionsprestellarcores_l1688}).
This shape is inverted toward the likely influence of  the O9V star $\sigma$~Sco  (see \citealp{2006MNRAS.368.1833N})  and is related to an interesting feature in the distribution of position angles for the Ophiuchus filaments. 
Figure~\ref{fig:filamentorientation} shows the distribution of median position angles for the sample of filaments  identified with \textsl{getfilaments}. 
It can be seen that this distribution  is bimodal, with a group of filaments roughly parallel to the large-scale streamers (cf. Sect.~2 and Fig.~\ref{oph_rgb}), and a group of filaments roughly perpendicular to the streamers, the latter group being associated with the L1688 main cloud. 
We compared the properties of starless cores embedded in parallel filamentswith those of cores embedded in perpendicular filaments, but did not find any clear evidence of differences between the two groups of cores.  

The two modes of filament orientations observed here are roughly centered around position angles P.A. $\sim \,$50 deg and  P.A. $\sim \,$130 deg (measured east of north), respectively, and may possibly be correlated with changes in the orientation of the magnetic field within the Ophiuchus cloud complex.
On large scales, {\it Planck} polarization data (\citealp{2016A&A...586A.138P}) show that the magnetic field orientation in the complex is mostly parallel to the streamers (P.A. $\sim \,$45 deg) and to the direction of the large-scale flows from the Sco~OB2 association. 
Close to L1688, however, the {\it Planck} data suggest that the magnetic field becomes almost perpendicular to the streamers, possibly as a result of large-scale compression by the Sco~OB2 flows.
While the angular resolution of {\it Planck} polarization data is insufficient to constrain the orientation of the field on small scales within the densest regions, optical and near-IR polarization results do suggest that  the magnetic field has a bimodal distribution of position angles within L1688, centered at P.A. $\sim \,$50 deg and  P.A. $\sim \,$150 deg (\citealp{1976AJ.....81..958V, 1988MNRAS.230..321S, 1994ApJ...424..208G}). 
This is strongly reminiscent of the two modes of filament orientations in Fig.~\ref{fig:filamentorientation}, pointing to a common origin.

\subsection{Filaments and cores}
The ubiquity of filaments in star-forming regions and the intimate link between prestellar cores and dense filaments is a crucial result of the {\it Herschel} HGBS survey (\citealp{2010A&A...518L.102A}). 
As prestellar cores are believed to be the direct precursors of protostars, this result suggests that dense molecular filaments may play a crucial role in the origin of protostars and  stars themselves. 

The high sensitivity and dynamic range of the {\it Herschel} images allow us to investigate this link between cores and filaments in the Ophiuchus complex, where there was no strong evidence of filamentary structures up to now, except for the large-scale streamers.

\begin{figure}[!htp]
\centering
\includegraphics[width=\hsize]{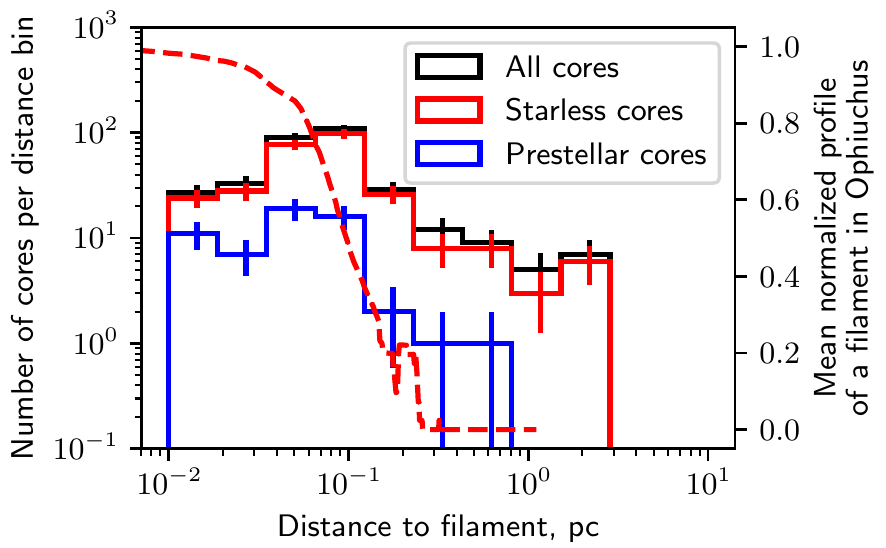}
\caption{Distribution of core-to-filament separations in the Ophiuchus cloud, for both prestellar cores and (unbound) starless cores. 
For each core, the projected separation to the nearest filament in the sample of filamentary structures extracted with DisPerSE (cf. Sect.~4.3) was considered. 
For comparison, the red dashed curve shows an example of a normalized, crest-averaged radial column density profile for a filament traced with DisPerSE in Ophiuchus.}
\label{fig:dtf}
\end{figure}

As observed with {\it Herschel} in other Gould Belt regions (\citealp{2010A&A...518L.102A, 2014prpl.conf...27A}), and indirectly hinted at  in previous studies of Ophiuchus (\citealp{1998A&A...336..150M}),  the prestellar cores of L1688 and L1689 tend to lie in elongated cloud structures or filaments.  
To quantify this trend, we carried out an analysis of the separations between dense cores and filaments, considering both unbound starless cores and prestellar cores. 
Figure~\ref{fig:dtf} shows that prestellar cores indeed lie within the $\sim\,$0.1\,pc inner portions of the densest filaments extracted with DisPerSE.\@ 
The population of unbound starless cores lies inside dense filaments, but also outside of them, displaying a higher median separation to the nearest filaments. 
This may have distinct explanations. 
First, unbound starless cores may in fact be split into two populations, with a subset of cores that will become prestellar cores by accumulating more mass on one hand, and transient or ``failed'' cores \citep[cf.][]{Ballesteros+2007} that will eventually disperse on the other hand. 
Second, our filament sample may be partly incomplete, especially in the low-density parts of the cloud where a significant fraction of unbound starless cores are found, and this may lead to a significant bias in the distribution of core-to-filament separations for unbound starless cores.
Nevertheless, the result that prestellar cores tend to lie closer to the crests of their nearest filaments than unbound starless cores provides an interesting clue suggesting that the star-forming activity of the cloud has not stopped, and that the lower-density material may possibly continue to form star-forming structures in the future, evolving toward another generation of prestellar cores and protostars by cloud contraction \citep[cf.][]{HuffStahler2007}.

\subsection{Core clustering and distribution of core separations}
\begin{figure*}[htpb]
\centering
\includegraphics[trim={0cm 1cm 0cm 2cm},clip,width=\hsize]{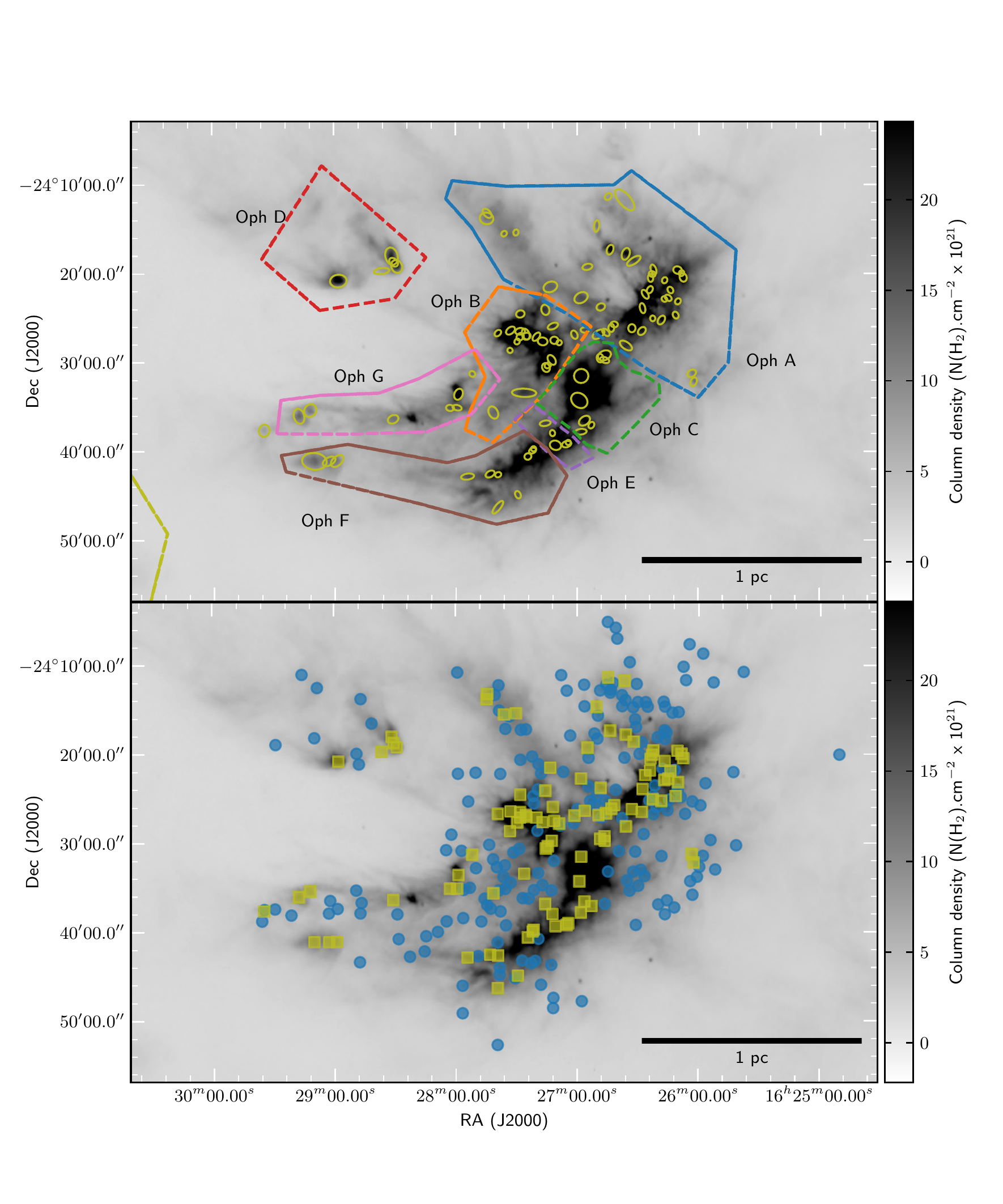}
\caption{{\bf (Top)} Spatial distribution of prestellar cores in L1688. 
The centroid positions, sizes, and position angles estimated with \textsl{getsources} for the prestellar and unbound cores in Table~\ref{tab_obs_cat} are represented by yellow ellipses overlaid on the {\it Herschel} high-resolution column density map derived from {\it Herschel} data (grayscale background image). Different colors are used for the separate dense clumps of L1688 (Oph A in blue, Oph B in orange, Oph C in green, Oph D in red, Oph E in purple, Oph F in brown, and Oph G in pink). 
{\bf (Bottom)} Compared spatial distribution of prestellar cores (yellow squares) and unbound cores (blue circles) overlaid on the high resolution column-density map.}
\label{fig:subregionsprestellarcores_l1688}
\end{figure*}
\begin{figure*}[!h]
\centering
\includegraphics[trim={0.8cm 0.5cm 2cm 1.3cm},clip,width=\hsize]{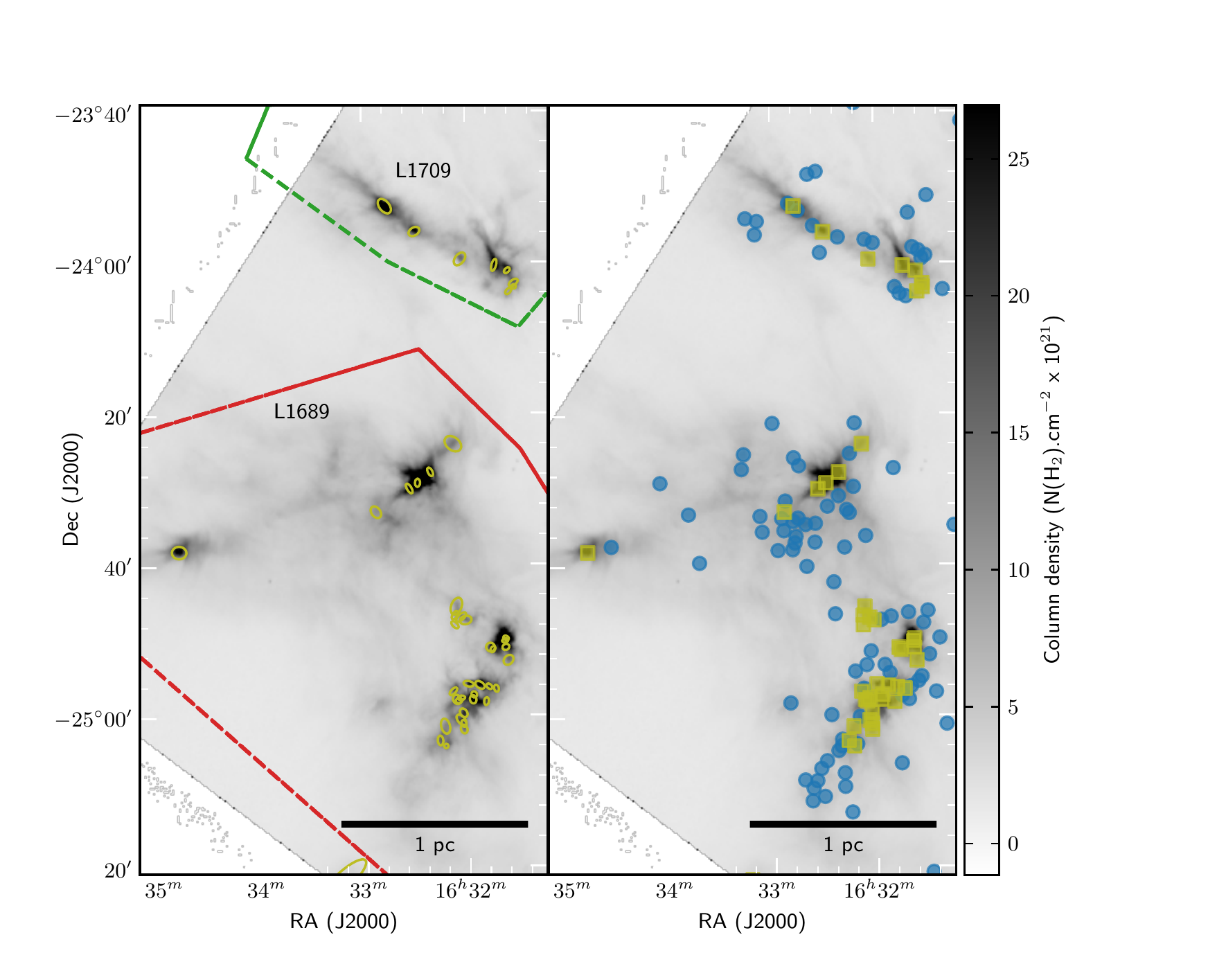}
\caption{{\bf (Left)} Spatial distribution of prestellar cores in L1689 and L1709.  
The centroid positions, sizes, and position angles estimated with \textsl{getsources} for the prestellar and unbound cores in Table~\ref{tab_obs_cat} are represented by yellow ellipses overlaid on the {\it Herschel} high-resolution column density map derived from {\it Herschel} data (grayscale background image). Different colors are used for the separate dense clumps of L1688 (L1689 in red, L1709 in green). 
{\bf (Right)} Compared spatial distribution of prestellar cores (yellow squares) and unbound cores (blue circles) overlaid on the high resolution column-density map.}
\label{fig:subregionsprestellarcores_l1689}
\end{figure*}

Figures~\ref{fig:subregionsprestellarcores_l1688} and \ref{fig:subregionsprestellarcores_l1689} show the spatial distribution of the candidate dense cores identified as explained in Sect.~\ref{sec:corext}, in relation with the distribution of cloud material as traced by our {\it Herschel} column density maps for the various subregions of the Ophiuchus complex. 
It can be seen that prestellar cores (yellow symbols) tend to be mostly clustered inside the densest regions, while unbound starless cores (blue symbols) are more widely spread out. 

Following the detection of DCO$^+$ gas clumps by \citet{1990ApJ...365..269L}, the L1688 cloud has been described as a collection of cold dense molecular clumps (called ``cores'' at the time), labeled with letters A to F. 
Based on our results, the most active regions for prestellar core growth appear to be Oph~A and Oph~B, while Oph~C, E, F harbor fewer prestellar cores but contain a higher number of Class~I protostars (\citealp{1989ApJ...340..823W}, \citealp{1994ApJ...420..837A}). 
No significant difference in dense core properties is observed between L1688 and L1689.
The sub clouds of L1688 (A, B, C, D, E, F, G) contain 40, 22, 8, 5, 5, 11, 7 prestellar cores,  respectively, and  79, 23, 11, 7, 3, 17, 11 unbound starless cores. 
Two prestellar cores and 60 unbound starless cores could not be assigned to any specific sub-region. 
Indeed, this historical breakdown of L1688 in sub-regions has been based on studies focusing on the densest gas in the cloud, and is therefore not well suited for classifying objects lying in lower-density areas. 
L1689 contains 35 prestellar cores and 79 unbound starless cores. 
L1709 contains 9 prestellar cores and 24 unbound starless cores. 

\begin{figure}[!h]
\centering
\includegraphics[width=\hsize]{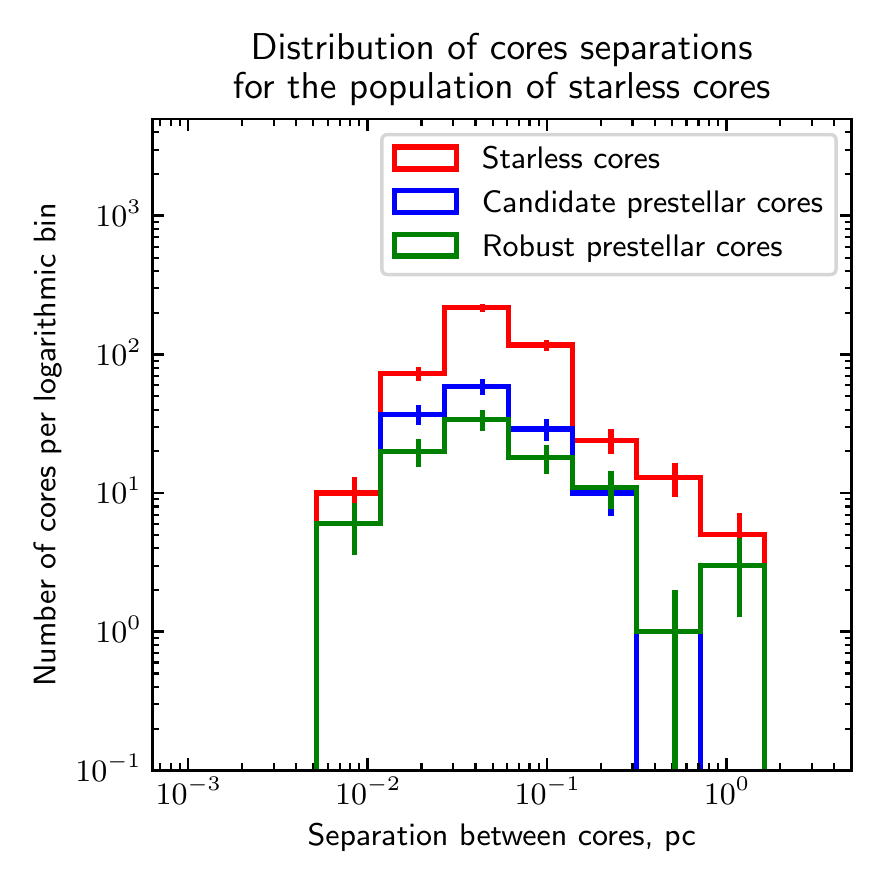}
\includegraphics[width=\hsize]{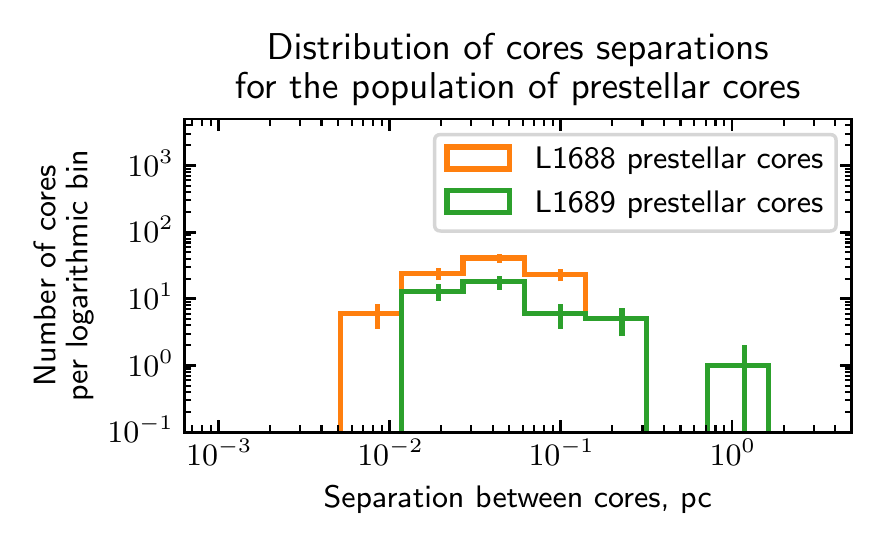}
\caption{{\bf (Top)} Distributions of nearest-neighbor separations for the populations of  starless and prestellar cores. 
{\bf (Bottom)} Comparison of the distributions of nearest-neighbor prestellar core separations in L1688 and L1689.}
\label{fig:coresep}
\end{figure}

Independently of background cloud structures and filaments, prestellar and protostellar cores tend to form in clusters, and therefore retain pristine properties of the parent molecular clouds in their distribution of separations. 
We can try to test this idea based on a simple statistical analysis of the distribution of core spacings.
In Fig.~\ref{fig:coresep}, we compare the distributions of nearest-neighbor core separations for the populations of prestellar cores and starless cores, in both L1688 and L1689. 
Even if the statistics are low, the different figures and modes of clustering show very little difference from one another, indicating that there may be only one underlying statistics in this sample, and probably a main physical explanation for the observed distribution of cores separations. 
Linked to molecular cloud structure, we observe a peak separation of $\sim 0.05\,$pc in both panels of Fig.~\ref{fig:coresep}, with a median nearest neighbor separation of $\sim 0.04\,$pc for prestellar cores. 
This is slightly smaller than, but comparable to, the characteristic inner width of {\it Herschel} filaments found by \citealp{Arzoumanian+2011,Arzoumanian+2019}\footnote{This is also similar to the widths of the N$_2$H$^+$ fiber-like structures observed by \citet{Hacar+2018} in the Orion integral filament region, but slightly smaller than the typical width of the fibers identified by \citet{Hacar+2013} in Taurus. }. 
More specifically, \citet{Arzoumanian+2019} recently measured a median inner width of 0.06$\pm$0.02 pc (FWHM) for 57 filaments in the same {\it Herschel} field of Ophiuchus.
It is also consistent with the typical Jeans length $c_s^2/(G \Sigma_{\rm cl})  \sim 0.04\,$~pc in cold dense molecular gas at a column density $N_{\rm H_2}^{\rm cl} \equiv \Sigma_{\rm cl} /(\mu_{H_2}\, m_H) \sim 10^{22}\, {\rm cm}^{-2} $. 
This result is physically not trivial, because \cite{1997ApJ...480..681I} found, through both analytical calculations and numerical simulations of  cylindrical cloud fragmentation, that the most unstable wavelength initiated in a gas cylinder would enhance the production of core structures with a characteristic separation of $\sim 4$ times the filament diameter, or about 0.4~pc for filaments of $\sim 0.1\,$pc width. 
While there is no clear explanation yet for this discrepancy in a seemingly filament-dominated mode of core formation, external factors such as large-scale compression or other physical effects like magnetic fields \citep{Nakamura+1993} or turbulence \citep{Clarke+2017} may possibly account for this result. 
\begin{table}[!h]
\caption{{\bf Characteristic extents of  various starless core populations}$^{\dag}$}         
\label{table:3}     
\centering                      
\begin{tabular}{c c c}      
\hline\hline               
Cloud   & Characteristic radius of       & Characteristic radius of   \\    
        &    bound-core population   & unbound-core population    \\  
	    &  (pc)                 &        (pc)             \\   
\hline                       
L1688       &  0.43 $\pm$ 0.02$^{\dag\dag}$    &  0.50 $\pm$ 0.02 \\
Oph A       &  0.20 $\pm$ 0.015   &  0.25 $\pm$ 0.015 \\
Oph B       &  0.10 $\pm$ 0.015   &  0.12 $\pm$ 0.015 \\
Oph C       &  0.05 $\pm$ 0.01    &  0.08 $\pm$ 0.01 \\
Oph D       &  0.03 $\pm$ 0.005   &  0.10 $\pm$ 0.02 \\
Oph E       &  0.02 $\pm$ 0.005   &  0.01 $\pm$ 0.002 \\
Oph F       &  0.12 $\pm$ 0.02    &  0.13 $\pm$ 0.015 \\
Oph G       &  0.09 $\pm$ 0.015   &  0.12 $\pm$ 0.015 \\
L1689       &  0.38 $\pm$ 0.035   &  0.58 $\pm$ 0.035 \\
L1709       &  0.20 $\pm$ 0.035   &  0.28 $\pm$ 0.03 \\
\hline                                   
\end{tabular}
\tablefoot{\tablefoottext{\dag}{Calculated from the equivalent diameter of the convex hull of each subset of cores.}
\tablefoottext{\dag\dag}{Uncertainties are calculated as proportional to $\frac{1}{2 \sqrt{N}}$ with $N$ the number of objects used to derive the convex hull of each subset of cores.}} 
\end{table}

Using the core positions derived from {\it Herschel} data  in each sub-cluster, we can also compare the areas occupied by the population of prestellar cores and the population of unbound starless cores.
To this purpose, we derived the convex hull areas corresponding to prestellar cores and unbound starless cores for each sub-cloud, and this led to a characteristic radius in each case given in Table~\ref{table:3}.
For most  sub clouds, the characteristic area occupied by prestellar cores is significantly smaller than the area occupied by unbound starless cores. 
This result is consistent with the overall correlation found in Sect.~4.5 between core mass and ambient cloud pressure (see bottom panel of Fig.~\ref{fig:regboundunboundcoresmasssize}) and shows us that prestellar cores are intimately related to the densest parts of the molecular cloud, while unbound starless cores are also found on the outskirts of the densest regions. 
This raises the issue of the physical nature of unbound starless cores. 
On one hand, many of these objects lie in portions of the cloud where the quantity of dense gas material may not be large enough for the cores to reach the critical Bonnor-Ebert mass even if they continue accumulating mass (cf. Fig.~\ref{fig:subregionsprestellarcores_l1688} and  Fig.~\ref{fig:subregionsprestellarcores_l1689}). 
Such unbound starless cores may be ``failed'' cores in the sense of \citet{Ballesteros+2007}. 
On the other hand, the whole L1688 cloud may be globally contracting toward its center of mass (see  \citealp{HuffStahler2007} and Sect.~5.6 below). 
If this is indeed the case, both the converging network of filaments and the starless cores of L1688 may become significantly denser on a typical timescale ranging from $\sim\,$$3\times t_{\rm ff}^{\rm cloud}\,$$ \sim \,$1.5\,Myr to the sound crossing time $R_{\rm cloud}/c_s \sim \,$3--4\,Myr, and a significant fraction of the unbound starless cores identified here with {\it Herschel} may evolve into self-gravitating prestellar cores (and later on protostars) on a similar timescale.

\subsection{Prestellar core formation efficiency in the Ophiuchus molecular cloud}

Section~4 described how we derived accurate mass estimates for both the cloud and the dense cores.
Both are important for characterizing the future star formation potential of the cloud.
Indeed, the ratio of the total mass in the form of cores to the total cloud mass is directly related to the efficiency of core formation process.
Furthermore, the fraction of cloud mass that is not involved in core formation at the present time can either form other cores in the future or remain as a quiescent background, not participating in the star formation process.

\begin{table*}[!h]
\caption{Main properties of the Ophiuchus subclouds}         
\label{table:4}     
\centering                       
\begin{tabular}{c c c c c c c c c c}      
\hline\hline               
Cloud       &  Total Cloud  & Cloud mass    & Number of    & M$_{\rm cores}$/M$_{\rm gas}$ &   M$_{\rm cores}$/M$_{\rm dense}$     & R$_{\rm gas}$ & n$_{\rm gas}^{\dag}$   & R$_{\rm dense}$ 
& n$_{\rm dense}$$^{\dag}$   \\  
            &  Mass   & A$_V$ $\ge$ 7 &  prestellar  &                       &                                &           &                       &             &                  \\    
            &      &               &  cores       &                       &                                   &           &  ($\times$10$^{3}$)   &             &  ($\times$10$^{3}$) \\    
	        & \msun    & \msun         &              &                       &                               & pc        & cm$^{-3}$             & pc          & cm$^{-3}$        \\    
\hline                     
L1688       & 980      & 415           & 99           & 5.1\%                 &  12\%                         &   1.0     & 3.4                   &    0.6      &  5.3            \\      
L1689+L1709 & 890      & 165           & 43           & 2.7\%                 &  14\%                         &   1.4     & 1.1                   &    0.4      &  10.5           \\
\hline                               
\end{tabular}
\tablefoot{\tablefoottext{\dag}{Average volume density calculated assuming spherical geometry for each subcloud.}} 
\end{table*}

As mentioned in  Sect.~\ref{sec:ophcloudmass} (see also Table~\ref{table:4}), the fractions of dense molecular gas in L1688 and L1689 are different, L1688 being denser, confirming the idea that L1688 may be a shock-compressed region (e.g \citealp{1986ApJ...306..142L}), where the accumulation of gas is enhanced by diverse external factors, while L1689 has significantly less dense gas. 
But the total gas masses in the two clouds are similar, and L1689 should be able to form, by large-scale contraction and fragmentation, enough clumps and cores to match the star formation activity of L1688, if it is characterized by the same core-formation efficiency. 
We find a dense molecular gas fraction, above  N$_{H_2}\,$$ \sim \,$$7 \times10^{21}\,$cm$^{-2}$ (or $A_{\rm V} \ga 7$), as high as 42\% in L1688 alone, compared to a dense gas fraction of only 18.5\% in L1689 and L1709.  In the whole field (Fig~\ref{fig:boundunboundfullmap}) we find a dense gas fraction of 15\%, similar to the values obtained in the {\it Herschel} studies of Aquila (24\%, \citealp{2015A&A...584A..91K}) and Orion B (13\%, \citealp{2019arXiv191004053K}), but significantly higher than the 1\% value found in the Musca and Pipe nebula clouds (cf. Table~1 of \citealp{Arzoumanian+2019}).
However, we observe very similar prestellar core formation efficiencies in the dense gas of L1688 and L1689+L1709, with 12\% and 14\% of dense gas mass in the form of prestellar cores, respectively.

\begin{figure}[!h]
\centering
\includegraphics[width=\hsize]{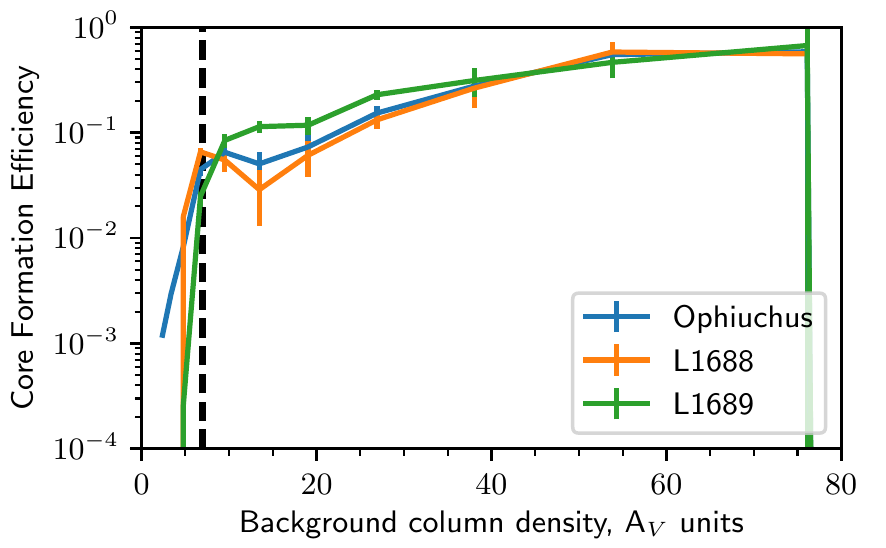}
\caption{Observed differential core formation efficiency (CFE) as a function of background column density expressed in A$_V$  units  (blue histograms with error bars) in the Ophiuchus molecular cloud, in blue for the whole cloud, in orange for L1688, and in green for L1689. The CFE if obtained by dividing the total mass of prestellar cores in a given column density bin by the cloud mass in this column density bin. The vertical dashed line marks a fiducial threshold at A$_V^{bg} \sim 7$. }
\label{fig:cfeffvscd}
\end{figure}
\begin{figure}[!h]
\centering
\includegraphics[width=\hsize]{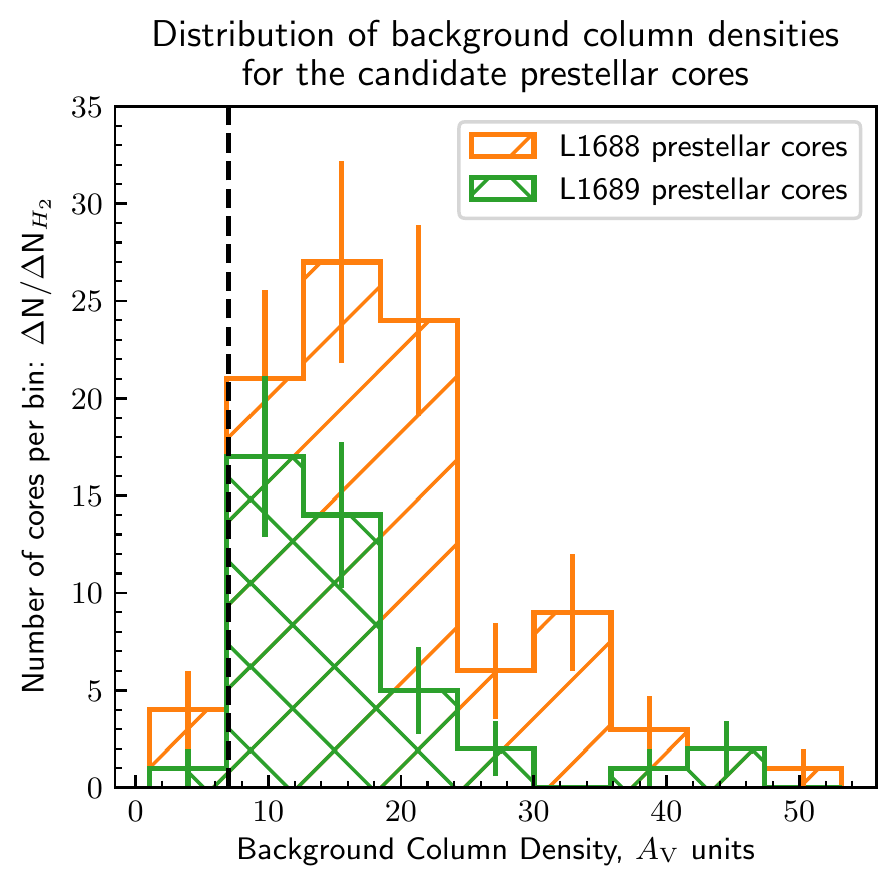}
\caption{Distribution of background column densities  for the L1688 (orange) and L1689 (green) prestellar cores. 
A clear peak in column density is visible, suggestive of a column-density threshold for prestellar core formation. 
However, this peak lies at a significantly higher column density  in L1688 (orange) and in L1689 (green) than in other Gould Belt regions such as Aquila \citep[cf.][]{2015A&A...584A..91K}.  
The vertical dashed line marks the same fiducial threshold at A$_V^{bg} \sim 7$ as in Fig.~\ref{fig:cfeffvscd}.}
\label{fig:regboundcoresbckcd}
\end{figure}

More precisely, following \citet{2015A&A...584A..91K}, we can define the differential prestellar core formation efficiency as a function of ``background'' cloud column density as follows: $ {\rm CFE_{obs}}(A_{\rm V}) = \Delta M_{\rm cores}(A_{\rm V})/\Delta M_{\rm cloud}(A_{\rm V}) $.
This is obtained by dividing the mass $\Delta M_{\rm cores}(A_{\rm V})$ of the candidate prestellar cores identified with {\it Herschel} in a given bin of background column densities (expressed in $A_{\rm V}$ units) by the total cloud mass $\Delta M_{\rm cloud}(A_{\rm V}) $ in the same bin. 
Figure~\ref{fig:cfeffvscd} shows a comparison of $ {\rm CFE_{obs}}$ vs. $A_{\rm V}$ for L1688, L1689, and Ophiuchus as a whole. It can be seen that all three  $ {\rm CFE_{obs}}(A_{\rm V})$ plots have similar shapes and exhibit similar values for $A_{\rm V} \ga 7$. 
Moreover, the differential $ {\rm CFE_{obs}}(A_{\rm V})$ plots level off above A$_V$ of 7-8 with typical values of $\sim \,$20\%.

In contrast, if we consider the total mass of molecular gas in each cloud, including low-density material, then L1688 appears significantly more efficient at forming prestellar cores, since the fraction of prestellar core mass to total gas mass is 5.1\%  there compared to only 2.7\% in L1689+L1709 (i.e., almost a factor of $\sim \,$2 lower than in L1688). 
This is most likely only due to the fact that L1688 contains more dense molecular gas ($A_{\rm V} > 7$) than L1689+L1709, however. These results are summarized in Table~\ref{table:4}.

A plot of the number of prestellar cores as a function of  background column density (Fig.~\ref{fig:regboundcoresbckcd}) shows a main peak around A$_V$ $\sim$10, as well as a secondary peak at higher extinction $A_{\rm V} \sim 30$ in L1688 and $A_{\rm V} \sim 40$ in L1689.
The secondary peak is related to the dense clumps  Oph~A and Oph~B in L1688, and to the immediate vicinity of IRAS16293 in L1689. 

Interestingly, while the physical meaning of this secondary peak remains difficult to assess, it appears to be related to features seen in the column density PDFs of the clouds (Fig.~\ref{fig:PDF}) and in the differential core formation efficiency plot  against background column density (Fig.~\ref{fig:cfeffvscd}). 
The main peak is related to a break in the column density PDF and a flattening in the differential $ {\rm CFE_{obs}}(A_{\rm V})$ plot. 
The secondary peak may result from a relative depletion of intermediate column density gas in L1688 because the corresponding material may be trapped  between the shells around S1 and HD147889 and compressed to higher column densities. 

Figure~\ref{fig:cfeffvscd}  and the main peaks in Fig.~\ref{fig:regboundcoresbckcd} are consistent with the presence of a ``threshold'' for the formation of prestellar cores at a fiducial A$_V$ value of $\sim\,$7--8, as seen in other HGBS regions \citep[e.g.,][]{2015A&A...584A..91K,2016MNRAS.459..342M}. 
This column density threshold should not be viewed as a strict step function but rather as a sharp transition between a regime of very low prestellar core formation efficiency ($< 1\% $) at $A_{\rm V} << 7$ and a regime of roughly constant (or slowly varying), moderately high  core formation efficiency ($\approx 20\% $) at $A_{\rm V} >> 7$ (cf. Fig.~\ref{fig:cfeffvscd}).
The details of the transition differ slightly in Ophiuchus compared to Aquila (\citealp{2015A&A...584A..91K}) and Taurus (\citealp{2016MNRAS.459..342M}), due to the presence of secondary peaks at higher column-densities in L1688 and L1689 which have not been seen in other regions. This slight difference may be due to the fact that the Ophiuchus cloud is the nearest region of the Gould Belt influenced by strong external compression, where prestellar cores can be detected (and resolved) in higher column-density areas than in previously studied HGBS regions. 
The overall transition seen in Fig.~\ref{fig:cfeffvscd} for L1688 and L1689 is nevertheless very similar to that observed in Aquila and Taurus. 
A physical interpretation of this transition in prestellar core formation efficiency around $A_{\rm V} \approx 7$ (or a cloud surface density $\Sigma_{\rm cl} \approx 150\, M_\odot$/pc$^2$) was proposed by \citet{2010A&A...518L.102A} in terms of the filamentary structure of the parent clouds.
Indeed, given the typical inner width $\sim 0.1\,$pc of molecular filaments \citep{Arzoumanian+2011,Arzoumanian+2019}, the transition at $\Sigma_{\rm cl} \approx 150\, M_\odot$/pc$^2$ closely corresponds to the critical mass per unit length $M_{\rm line, crit} = 2\, c_s^2/G \sim 16\, M_\odot$/pc (see also \citealp{2014prpl.conf...27A}), around which nearly isothermal gas filaments at $T \sim 10\,$K are known to fragment efficiently (Inutsuka \& Miyama 1997).

\subsection{Starless and prestellar core mass functions}
\begin{figure}[!htp]
\includegraphics[width=\hsize]{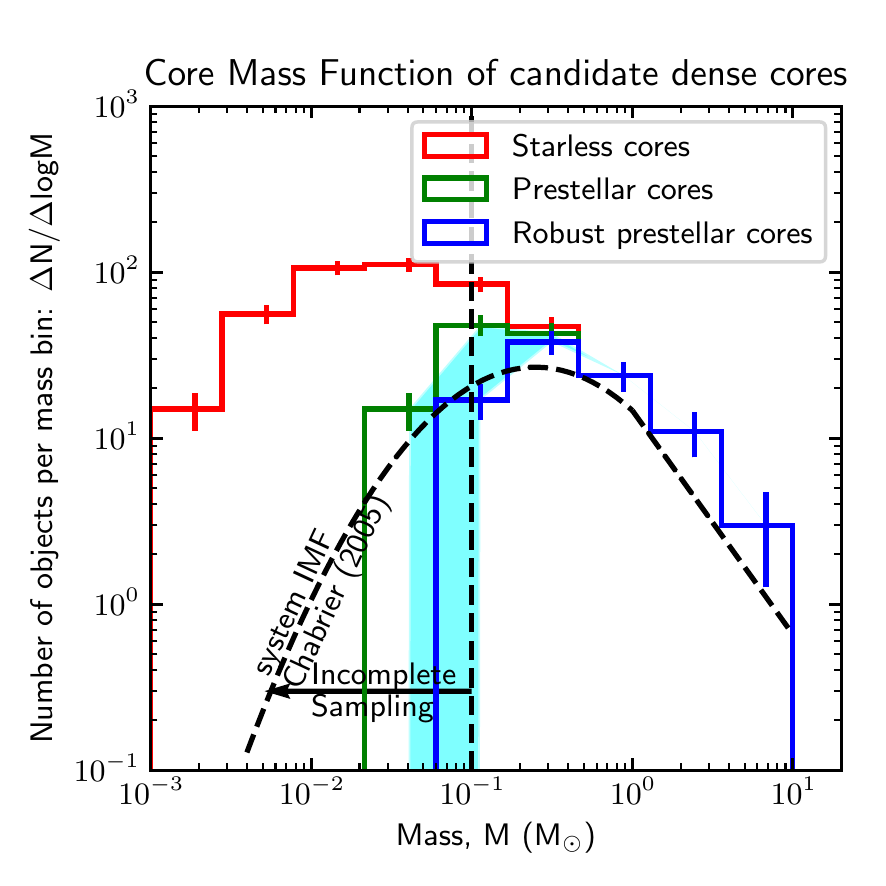}
\caption{Mass distribution of robust prestellar cores (blue), candidate prestellar cores (green), and unbound starless cores (red). 
The cyan area marks the difference between 
robust and candidate prestellar cores. 
To left of the vertical dashed line, the sample of prestellar cores is less than 80\% complete according to Monte-Carlo simulations 
(see Sect.~\ref{sec:completeness} and Appendix~B). 
For comparison, the black dashed curve shows the system IMF \citep{2005ASSL..327...41C}.}
\label{bound_unbound_cores_cmf}
\end{figure}

The mass distribution of our sample of 144 candidate prestellar cores provides an estimate of the prestellar core mass function (CMF) in the Ophiuchus cloud (Fig.~\ref{bound_unbound_cores_cmf}). 

\begin{figure*}[!h]
\centering
\includegraphics[trim={0cm 0cm 0cm 1cm},clip,width=\hsize]{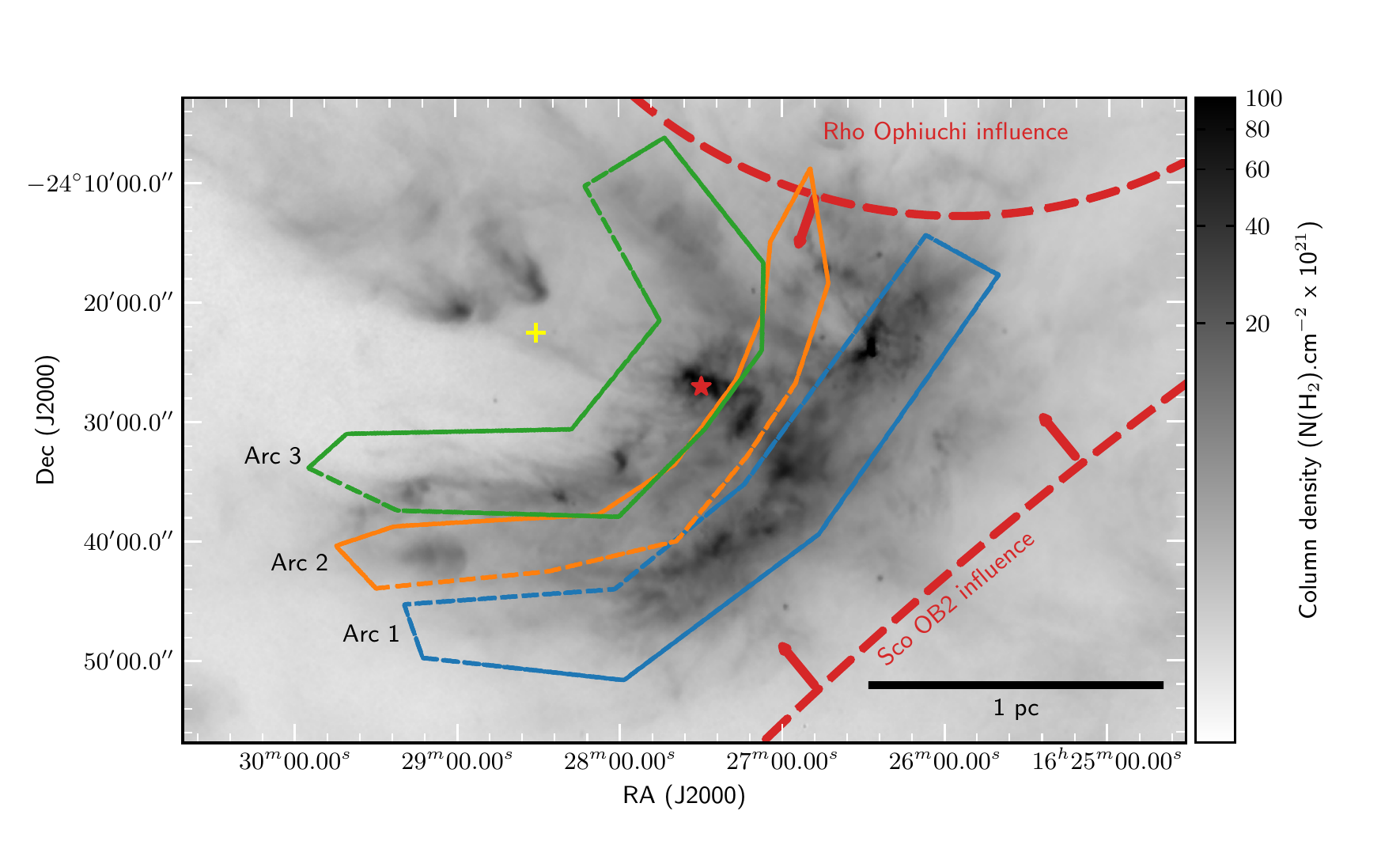}
\caption{Sketch of  three putative compression fronts in dashed blue (Arc 1), dashed yellow (Arc 2), and dashed green (Arc 3), 
corresponding to density enhancements, overlaid 
on the {\it Herschel} high-resolution column density map of L1688 (grayscale).  All three compression arc-like structures 
point roughly toward the direction of the O9V star $\sigma$ Sco (cf. Fig.~\ref{oph_rgb}), and are possibly skewed by the influence 
of the bubble around the star $\rho\,$Ophiuchi.
The yellow cross marks the rough center of the three arcs and the red star indicates the center of mass of L1688.}
\label{fig:sfarcs}
\end{figure*}

The prestellar CMF in Ophiuchus shows an apparent peak at about 0.3 \msun, which is very similar to the CMFs derived from HGBS data in Taurus/L1495 \citep{2016MNRAS.459..342M} and is also consistent within uncertainties with previous results in Aquila \citep{2015A&A...584A..91K} where the prestellar CMF seemed to peak around 0.4--0.6\,\msun. 
It should be stressed that the prestellar core masses derived here are likely underestimated  by a factor of $\sim 2$ on average, due to the extraction process and the effect of temperature gradients within the cores.
This has been evaluated based on the Monte-Carlo simulations described in Appendix~B where  synthetic prestellar cores were injected injecting  in the  {\it Herschel} maps.

The prestellar CMF also has a direct relationship with the IMF, emphasizing the importance of core formation studies in molecular clouds \citep[cf.][]{1998A&A...336..150M}. 
The close resemblance of the prestellar CMF to the stellar system IMF (\citealp{2005ASSL..327...41C}) is clearly confirmed in Ophiuchus (cf. Fig.~\ref{bound_unbound_cores_cmf}), like in Aquila \citep{2015A&A...584A..91K} and Taurus \citep{2016MNRAS.459..342M}.
However, while the total number of prestellar cores identified here (144) is significantly higher than in previous studies of Ophiuchus \citep[e.g][]{1998A&A...336..150M, Johnstone+2000, Stanke+2006, 2015MNRAS.450.1094P}, the core statistics in this cloud are still not high enough to allow robust, final conclusions to be drawn about the prestellar CMF. 

\subsection{Long-distance external effects and positive feedback}

\begin{figure*}[!h]
\centering
\includegraphics[trim={0cm 0.3cm 0cm 0cm},clip,width=\hsize]{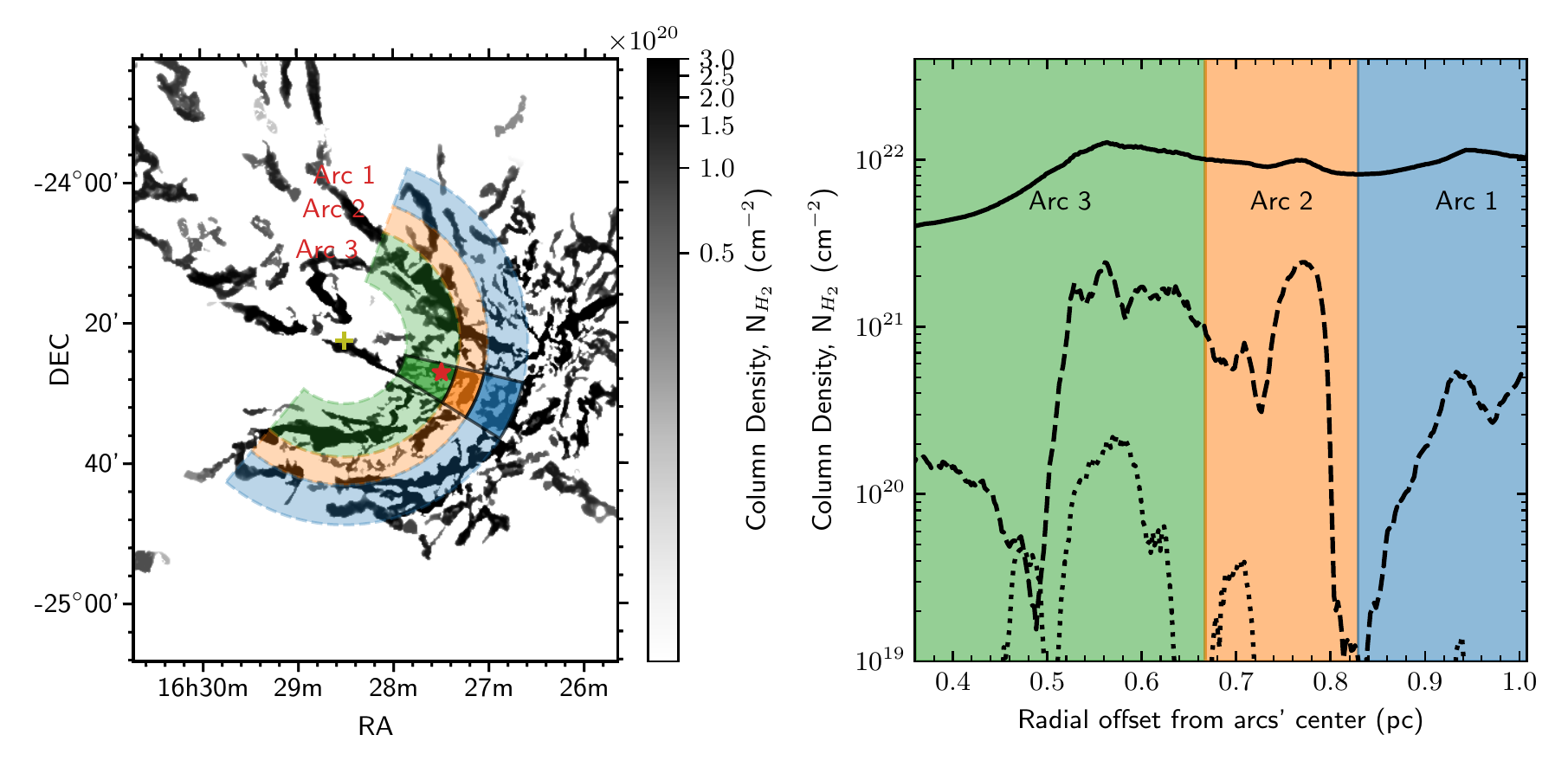}
\caption{{\bf (Left)} Filtered column density map of L1688 where concentric annuli associated with the shaded ranges are overplotted, each annulus marking the presence of one arc-like over-compression as seen in the right panel. 
The yellow cross marks the center of the concentric annuli and the red star indicates the center of mass of L1688.
{\bf (Right)} Radial column density profile of the L1688 cloud, circularly averaged about a center close to Oph~D.
The shaded ranges highlight the three compression fronts seen in the column density image and associated with the three arcs in the left panel. 
The solid curve shows the profile obtained from an azimuthal average of the unfiltered column density map shown in Fig~\ref{fig:sfarcs} over the whole 190$^\circ$ of the shaded sectors in the left panel.
The dashed curve shows the profile obtained from an azimuthal average of the filtered column density map over the same rangeof the shaded sectors in the left panel.
The dotted line shows the profile averaged over the angles excluding the part around Oph~B, outside of the two black solid lines in the left panel, with evidence of an overdensity highlighting the second arc (Arc 2).  
}
\label{fig:sfarcsfit}
\end{figure*}

There appears to be an imprint of at least three compression fronts in the {\it Herschel} column density map of L1688 (see Fig.~\ref{fig:sfarcs}), which can be confirmed by performing a circular average of the cloud emission about a central point located close to Oph~D.
These compression fronts are detected in the column density profile of the cloud, with three consecutive local peaks in the cloud profile as a function of radius from a center near Oph~D (see Fig.~\ref{fig:sfarcsfit}).
The two most prominent peaks correspond first to an arc-like area crossing Oph~B2 (named Arc~3 in Fig.~\ref{fig:sfarcs}), and second to an arc-like area (named Arc~1) following the well-known dense C$^{18}$O ridge  of L1688 \citep{Wilking+1983}. 
A third, shallower peak seen in the circularly-averaged column density profile of Fig.~\ref{fig:sfarcsfit} corresponds to the arc-like area named Arc~2 in Fig.~\ref{fig:sfarcs}, which includes Oph B1. 
A hint of the presence of this arc was already visible in the form of a weak filamentary structure in the ground-based 1.2\,mm continuum map of  \citet[][see their Fig.~1]{Stanke+2006}. 

The three consecutive arc-like structures Arcs 1--3 may originate from compression by large-scale shockwaves propagating through the cloud from the Sco~OB2 association \citep{1986ApJ...306..142L, 1998A&A...336..150M}. 
There are at least two possibilities for the origin of these multiple arc-like features.
First,  the presence of multiple compression fronts is reminiscent of the theoretical scenario proposed by \cite{2015A&A...580A..49I} for the formation of molecular clouds based on multiple shockwave compressions of interstellar atomic gas. 
Second, the multiple compression fronts may trace the interaction of a single shockwave with multiple preexisting structures within the L1688 cloud. In the latter case, however, we would expect to observe an age gradient from northeast to southwest with younger YSOs associated with Arc~3 (compressed last) and more evolved YSOs associated with Arc~1 (compressed first). 
In contrast, the highest concentration of Class~I \citep{1989ApJ...340..823W,2001A&A...372..173B} and even Class~0 \citep{1993ApJ...406..122A} objects is found in the C$^{18}$O ridge (Arc~1), while a number of relatively evolved Class~II/III objects and X-ray emitting weak-line T Tauri stars are observed close to Arc~3 \citep{Casanova+1995}. 
For this reason, we favor the former interpretation of multiple compression events and shockwaves.  
The interaction of each shockwave with an overdensity cloud structure created in a previous compression event can generate a bow-shape or U-shape feature similar to the arcs  seen in Fig.~\ref{fig:sfarcs} (cf. Fig.~1 of \citealp{Inoue+2013}). 
The cometary shape of the Oph~D dense clumps is also consistent with this scenario.  

The Ophiuchus complex, and the L1688 cloud in particular, are known to be influenced and possibly shock-compressed by an expanding shell of HI gas powered by stellar winds  and supernova explosions in the Sco~OB2  association, including the effect of the neighboring massive (O9V) star $\sigma$ Sco 
\citep{1986ApJ...306..142L, deGeus1992, 2006MNRAS.368.1833N}. 
Based on Fig.~\ref{fig:sfarcs}, we suggest that the main filaments of L1688 may result from three consecutive shock compressions.
The region studied here is in fact enclosed between Antar\`es ($\alpha$ Sco) and $\sigma$ Sco (cf. Fig.~\ref{oph_rgb}). 
The feedback influences of these stars may be significant as suggested by the positions of these OB stars relative to the three arcs identified here
(Fig.~\ref{oph_rgb}, Fig.~\ref{fig:sfarcs}). 

Given the estimated age of $\la 2\,$Myr for the L1688 embedded cluster \citep{2001A&A...372..173B} and the different number ratios of prestellar to unbound starless cores observed in the three arcs (see Table~5), it seems reasonable to assume a typical age difference of only $\la 1$~Myr between these arc-like compression fronts, similar to the typical timescale adopted by \citet{1977ApJ...218..148M} and \citet{2015A&A...580A..49I} for the dynamics of the Galactic interstellar medium resulting from supernova explosions and expanding HII regions.
The large-scale HI shell propagating into the Ophiuchus cloud complex expands at a velocity of about 10--15\, km\, s$^{-1}$ according to \citet{deGeus1992}, but the velocity of the transmitted shock within the  L1688 dense cloud  may be only $\la 1\, $km\, s$^{-1}$ based on the range of line-of-sight velocities observed in optically thin tracers of molecular gas \citep{1989ApJ...338..925L,1986ApJ...306..142L}.
This kind of low-velocity shock would not destroy the cloud \citep{1986ApJ...306..142L, 1981AJ.....86..885H, Foster+1996} and may ideally permit the formation of gravitationally-unstable condensations such as the candidate pre-brown dwarf Oph B-11 \citep{Greaves+2003, Andre+2012} in regions of the cloud with relatively low initial density.

\begin{table}[!h]
\caption{Ratios of various types of cores in the three arcs}
\label{table:5}     
\centering                        
\begin{tabular}{c c c c}       
\hline\hline               
Arc    & Number ratio of     &     Mass ratio      & Number ratio of     \\   
           &   bound to    &       bound to      & protostars$^{\dag}$ to     \\   
	    &  unbound cores     &        unbound cores  & unbound cores   \\   
\hline                      
Arc 1       &  1.2    &  21.8 & 0.5 \\
Arc 2       &  0.6    &  7.4  & 0.1 \\
Arc 3       &  0.7    &  12.2 & 0.13 \\
\hline
\end{tabular}
\tablefoot{\tablefoottext{\dag}{Comparison between the number of Class I protostars detected with {\it Spitzer} in \citet{2013AJ....145...94D} 
and the number of unbound starless cores in the present paper.}} 
\end{table}

While large-scale flows may be shaping the whole complex, looking for local hints of their effect on the small-scale structure of L1688 is interesting.
The spatial distribution of the dense cores identified in this paper shows global relationships, such as overall spreads and mean separations, but also contains information about the structure of the molecular cloud at a given time.
Also relevant are the substructures identified in the column density map of the complex.
Historically divided into clumps \citep[e.g.,][]{1990ApJ...365..269L}, the Ophiuchus complex had never been observed  with high enough sensitivity and dynamic range up to now to unveil the underlying filamentary texture of the cloud. 
While many clumps (A to G) were apparent in this region,  the {\it Herschel} observations now allows us to distinguish three nearly concentric arcs in the main cloud (Fig.~\ref{fig:sfarcs}).
This leads us to propose another classification of YSOs, prestellar cores, and unbound starless cores in L1688, not divided into dense clumps as in \citet{1990ApJ...365..269L}but divided into three concentric arcs Fig.~\ref{fig:sfarcs}. 
Simple statistics about the ratios of bound to unbound cores in the three arcs are summarized in 
Table~\ref{table:5}. 
We notice that the second or central arc (Arc 2) seen in the azimuthal average of L1688 (Fig.~\ref{fig:sfarcsfit}) is significantly more populated with unbound starless cores, in the middle of the cloud. 
In contrast, Arc 1 and Arc 3 are seemingly more evolved than Arc 2 with more YSOs and prestellar cores inside of them. 
With signs of global contraction in this cloud from self-absorbed CO profiles skewed to the blue (\citealp{1974ApJ...189L.135E}, \citealp{1980ApJ...238..620L}), it is tempting to speculate that some of the unbound starless cores observed within Arc 2 will be further compressed in the future and accumulate more mass, eventually leading to the formation of new prestellar cores and a new generation of stars in L1688. 

\section{Conclusions}

Parallel-mode (SPIRE and PACS) maps obtained as part of the {\it Herschel} Gould Belt survey were used to carry out an extensive study of the population of dense cores in the Ophiuchus molecular cloud (L1688, L1689, L1709). 
We summarize our main results as follows:

\begin{enumerate}
\item Using the \textit{Herschel} multi-wavelength data, we derived a high-resolution ($18.2\arcsec $) column density map of the Ophiuchus complex (Figs.~\ref{oph_rgb}  \& \ref{fig:boundunboundfullmap}). 
This map clearly reveals for the first time the presence of  filaments inside the main cloud, L1688, which was not previously known to be a filamentary region. L1688 appears to be structured in the form of a converging network of filaments pointing toward its center of mass near Oph~B2 (see Fig.~\ref{fig:boundunboundfullfilmap}--top).
The filaments have a bimodal distribution of orientations (cf. Fig.~\ref{fig:filamentorientation}), either roughly parallel or roughly perpendicular to the direction of the large-scale streamers of the complex.
\item We identified a total of 144 candidate prestellar cores, including 93 robust prestellar cores, in L1688 and L1689/L1709, using the \textsl{getsources} source extraction algorithm after a careful selection and inspection of each source. 
In addition, 315 unbound starless cores and 54 protostellar cores were also detected. 
The main properties (e.g., radius, mass, temperature, density)  of all of these cores were characterized (cf. online Table~A.2) thanks to a good sampling of the SEDs in the far-infrared and submillimeter range. 
\item The mass distribution of the prestellar cores closely resembles the IMF of stellar systems (Fig.~\ref{bound_unbound_cores_cmf}) confirming earlier results.
\item The prestellar cores lie preferentially within dense filaments and are mostly found very close ($< 0.1\,$pc) to the crests of their parent filaments (Fig.~\ref{fig:dtf}).  They are also significantly more concentrated in the densest parts of the complex, occupy a significantly smaller area of the total field (Figs.~\ref{fig:subregionsprestellarcores_l1688} \& \ref{fig:subregionsprestellarcores_l1689}), and tend to lie in regions of higher ambient cloud pressure (Fig.~\ref{fig:regboundunboundcoresmasssize}) than the unbound starless cores.
\item The median nearest neighbor separation between prestellar cores is $\sim\,$0.04\,pc, which is comparable to the typical Jeans length in cold dense molecular gas at $N_{\rm H_2} \sim 10^{22}\, {\rm cm}^{-2} $. The typical projected core separation is also comparable to the median inner filament width $\sim 0.06\,$ pc measured by \citet{Arzoumanian+2019} in Ophiuchus but significantly shorter than the characteristic core spacing of $\sim 4$ times the filament width ($\sim 0.24\,$ pc here) expected from standard cylinder fragmentation models. This difference may result from the effects of magnetic fields and turbulence, or the fact that the Ophiuchus filaments are not isolated systems.
\item We confirmed the presence of a ``threshold'' or sharp transition in column density around a fiducial A$_V\,$ value of $\sim\,$7 for the formation of prestellar cores (Figs.~\ref{fig:cfeffvscd}  \& \ref{fig:regboundcoresbckcd}). 
Moreover, by comparing our results in  L1688 and L1689, we found clear evidence that only a fraction of the total mass budget of the complex is directly participating in the star formation process.
Indeed, L1689 is less efficient at forming stars than L1688, but shares almost the same total mass of molecular gas. 
When only considering the dense molecular gas above A$_V$$\,\sim \,$7,  L1689 and L1688 appear to be equally efficient at forming dense cores.
\item The prestellar core formation threshold is strongly correlated to the presence of dense filaments in the Ophiuchus complex. 
Most of the dense gas material in the cloud and most of the prestellar cores are distributed in dense filaments within several clumps. 
Furthermore, the majority of prestellar cores lie within the $\sim\,$0.1\,pc inner portions of their parent filaments (Fig.~\ref{fig:dtf}). 
The close connection between prestellar cores and filaments is especially remarkable here in Ophiuchus, as there was no direct detection of filamentary structures up to now in the main region, L1688.
\item Three shock compression fronts can be detected in the column density map of the L1688 cloud, locally enhancing the quantity of dense molecular gas and supercritical filaments, hence the formation of prestellar cores. 
\end{enumerate}

\begin{acknowledgements}

We thank Shu-ichiro Inutsuka for insightful discussions on cloud compression and the referee, Thomas Stanke, for useful comments which improved the clarity of the paper. 
SPIRE has been developed by a consortium of institutes led by Cardiff Univ. (UK) 
and including: Univ. Lethbridge (Canada); NAOC (China); CEA, LAM (France); 
IFSI, Univ. Padua (Italy); IAC (Spain); Stockholm Observatory (Sweden); 
Imperial College London, RAL, UCL-MSSL, UKATC, Univ. Sussex (UK); 
and Caltech, JPL, NHSC, Univ. Colorado (USA). This development has been 
supported by national funding agencies: CSA (Canada); NAOC (China); 
CEA, CNES, CNRS (France); ASI (Italy); MCINN (Spain); SNSB (Sweden); 
STFC, UKSA (UK); and NASA (USA). 
PACS has been developed by a consortium of institutes led by MPE
(Germany) and including UVIE (Austria); KUL, CSL, IMEC (Belgium); CEA,
OAMP (France); MPIA (Germany); IFSI, OAP/AOT, OAA/CAISMI, LENS, SISSA
(Italy); IAC (Spain). This development has been supported by the funding
agencies BMVIT (Austria), ESA-PRODEX (Belgium), CEA/CNES (France),
DLR (Germany), ASI (Italy), and CICT/MCT (Spain). 
This work has received support from the European Research Council 
under the European Union's Seventh Framework Programme 
(ERC Advanced Grant Agreement no. 291294 --  `ORISTARS') 
and from the French National Research Agency (Grant no. ANR--11--BS56--0010 -- `STARFICH'). 
We also acknowledge support from the French national programs of CNRS/INSU on stellar and ISM physics (PNPS and PCMI). 
YS was supported by NAOJ ALMA Scientific Research Grant (2017-04A). 
N.S. acknowledges support by the French ANR and the German DFG through 
the project "GENESIS" (ANR-16-CE92-0035-01/DFG1591/2-1).
PP and DA acknowledge support by FCT/MCTES through national funds (PIDDAC) by this grant UID/FIS/04434/2019. 
PP also acknowledges support from the fellowship SFRH/BPD/110176/2015 funded by FCT (Portugal) and POPH/FSE (EC).
This research made use of Astropy,\footnote{http://www.astropy.org} a community-developed core Python package for Astronomy \citep{astropy:2013, astropy:2018}.
\end{acknowledgements}

\bibliographystyle{aa} 
\bibliography{bibliography.bib} 

\newpage
\appendix

\section{A catalog of dense cores identified with {\it Herschel} in the {\bf Ophiuchus cloud}}\label{sec:appendix_catalog}

Based on our {\it Herschel} SPIRE/PACS parallel-mode imaging survey of the Ophiuchus molecular cloud, we identified a total of 513 dense cores, including 459 starless cores (144 of them being self-gravitating and prestellar in nature), and 54 protostellar cores.
The master catalog listing the observed properties of all of these {\it Herschel} cores is available in online Table~A.1 (see below for an illustrative portion of this table and http://gouldbelt-herschel.cea.fr/archives for the full table).
The derived properties (physical radius, mass, SED dust temperature, peak column density at the resolution of the 500~$\mu$m data, average column density, peak volume density, and average density) of the same cores are provided in online Table~A.2 (see also below for an illustrative portion and http://gouldbelt-herschel.cea.fr/archives for the full table).

\clearpage

\begin{sidewaystable*}[htbp]\tiny\setlength{\tabcolsep}{2.5pt}
\caption{Catalog of  dense cores identified in the HGBS maps of the Ophiuchus complex (template, full catalog only provided online).} 
\label{tab_obs_cat}
{\renewcommand{\arraystretch}{0.5}
\begin{tabular}{|r|c|c c c c c c c c c c} 
\toprule[1.0pt]\toprule[0.5pt]  
 rNO      & Core name         &  RA$_{\rm 2000}$ &  Dec$_{\rm 2000}$        & Sig$_{\rm 070}$ &  $S^{\rm peak}_{\rm 070}$ &  $S^{\rm peak}_{\rm 070}$/$S_{\rm bg}$ &  $S^{\rm conv,500}_{\rm 070}$ &  $S^{\rm tot}_{\rm 070}$ &  FWHM$^{\rm a}_{\rm 070}$ &  FWHM$^{\rm b}_{\rm 070}$ &  PA$_{\rm 070}$   \\ 
          & HGBS\_J*          &  (h m s)         &  (\degr~\arcmin~\arcsec) &                 & (Jy/beam)                 &                                        & (Jy/beam$_{\rm 500}$)         &  (Jy)                    &  (\arcsec)                &  (\arcsec)                &  (\degr)          \\         
 (1)      & (2)               &  (3)             &  (4)                     &  (5)            &      (6) ~ $\pm$ ~ (7)    &  (8)                                   & (9)                           &    (10) ~ $\pm$ ~ (11)   &  (12)                     &  (13)                     &  (14)             \\         
\toprule[0.8pt] 
$\cdots$  &                   &                  &                          &                 &                           &                                        &                               &                          &                           &                           &                   \\
  89      & 162626.6-242431   & 16:26:26.68      &  --24:24:31.0            &   706.8         &    1.29e+01 ~ 2.6e+00     &   4.08                                 &   2.09e+01                    &   2.81e+01  ~ 4.9e+00    &  13                       &  8                        &	 159              \\
$\cdots$  &                   &                  &                          &                 &                           &                                        &                               &                          &                           &                           &                   \\
  91      & 162627.6-242359   & 16:26:27.65      &  --24:23:59.3            &     0.0         &    9.69e-01 ~ 1.5e+00     &   0.09                                 &   9.33e+00                    &   3.51e+01  ~ 3.6e+00    &  57                       &  24                       &	 143              \\
$\cdots$  &                   &                  &                          &                 &                           &                                        &                               &                          &                           &                           &                   \\
  458     & 163223.1-242836   & 16:32:23.16      &  --24:28:36.5            & 11910.0         &    1.14e+02 ~ 8.1e-01     &   4.71                                 &   2.72e+02                    &   3.01e+02  ~ 9.5e-01    &  21                       &  8                        &  202              \\
$\cdots$  &                   &                  &                          &                 &                           &                                        &                               &                          &                           &                           &                   \\
\midrule[0.5pt]
\end{tabular}
}
\scalebox{1.2}{$\sim$}
\vspace{0.2cm}

\scalebox{1.2}{$\sim$}
{\renewcommand{\arraystretch}{0.5}
\begin{tabular}{c c c c c c c c c  c c c c c c c} 
\toprule[1.0pt]\toprule[0.5pt]  
 Sig$_{\rm 160}$ &  $S^{\rm peak}_{\rm 160}$    &  $S^{\rm peak}_{\rm 160}$/$S_{\rm bg}$ &  $S^{\rm conv,500}_{\rm 160}$ &  $S^{\rm tot}_{\rm 160}$ &  FWHM$^{\rm a}_{\rm 160}$ &  FWHM$^{\rm b}_{\rm 160}$ &  PA$_{\rm 160}$  &  Sig$_{\rm 250}$ &  $S^{\rm peak}_{\rm 250}$ &  $S^{\rm peak}_{\rm 250}$/$S_{\rm bg}$ &  $S^{\rm conv,500}_{\rm 250}$ &  $S^{\rm tot}_{\rm 250}$ &  FWHM$^{\rm a}_{\rm 250}$ &  FWHM$^{\rm b}_{\rm 250}$ &  PA$_{\rm 250}$  \\ 
                 &  (Jy/beam)                   &                                        & (Jy/beam$_{\rm 500}$)         &  (Jy)		    &  (\arcsec)		&  (\arcsec)		    &  (\degr)         &		  & (Jy/beam)		      & 				       & (Jy/beam$_{\rm 500}$)         &  (Jy)  		  &  (\arcsec)  	      &  (\arcsec)		  &  (\degr)	     \\       
  (15)           &      (16) ~ $\pm$ ~ (17)     &   (18)                                 &  (19)   			 &    (20) ~ $\pm$ ~ (21)   &  (22)                     & (23)                      & (24)             & (25)             &    (26) ~ $\pm$ ~ (27)    &  (28)                                  &  (29)                         &   (30) ~ $\pm$ ~ (31)    &  (32)                     &  (33)                     &  (34)            \\        
\toprule[0.8pt]
$\cdots$         &                              &                                        &                               &			    &				&			    &		       &		  &			      & 				       &                               &			  &			      & 			  &		     \\
  2316.0         &  6.35e+01 ~ 8.7e+00          &  2.33				         &  9.36e+01			 &  1.40e+02 ~ 1.3e+01      &  15			&  12		            &  142	       &   2841.0	          &  9.41e+01 ~ 1.2e+01       &    2.28				       &  9.40e+01		       &  1.17e+02 ~ 1.2e+01      &  19			      &  18		          &   93             \\
$\cdots$         &                              &                                        &                               &			    &				&			    &		       &		  &			      & 				       &                               &			  &			      & 			  &		     \\
  2531.0         &  6.53e+01 ~ 1.3e+01          &  1.30				         &  2.12e+02			 &  3.46e+02 ~ 2.4e+01      &  32		        &  16		            &  59	       &   3425.0	          &  1.57e+02 ~ 1.1e+01       &    2.36				       &  2.23e+02		       &  3.20e+02 ~ 1.3e+01      &  35			      &  18		          &   55             \\
$\cdots$         &                              &                                        &                               &			    &				&			    &		       &		  &			      & 				       &                               &			  &			      & 			  &		     \\
  21580.0        &  4.58e+02 ~ 3.0e+00          &  8.00				         &  7.44e+02			 &  8.02e+02 ~ 3.1e+00      &  17			&  12		            &  186	       &   16190.0	  &  4.78e+02 ~ 3.9e+00       &    8.46				       &  5.06e+02		       &  5.19e+02 ~ 3.9e+00      &  21			      &  18		          &  186             \\
$\cdots$         &                              &                                        &                               &			    &				&			    &		       &		  &			      & 				       &                               &			  &			      & 			  &		     \\
\midrule[0.5pt]
\end{tabular}
}
\scalebox{1.2}{$\sim$}
\vspace{0.2cm}

\scalebox{1.2}{$\sim$}
{\renewcommand{\arraystretch}{0.5}
\begin{tabular}{c c c c c c c c c c c c c c c c}  
\toprule[1.0pt]\toprule[0.5pt] 
Sig$_{\rm 350}$  &  $S^{\rm peak}_{\rm 350}$ &  $S^{\rm peak}_{\rm 350}$/$S_{\rm bg}$ &  $S^{\rm conv,500}_{\rm 350}$ &  $S^{\rm tot}_{\rm 350}$  &  FWHM$^{\rm a}_{\rm 350}$ &  FWHM$^{\rm b}_{\rm 350}$ &  PA$_{\rm 350}$   &  Sig$_{\rm 500}$ &  $S^{\rm peak}_{\rm 500}$ &  $S^{\rm peak}_{\rm 500}$/$S_{\rm bg}$ &  $S^{\rm tot}_{\rm 500}$ &  FWHM$^{\rm a}_{\rm 500}$ &  FWHM$^{\rm b}_{\rm 500}$  &  PA$_{\rm 500}$   \\
                 &  (Jy/beam)                &                                        & (Jy/beam$_{\rm 500}$)         &  (Jy)			  &  (\arcsec)  	      &  (\arcsec)		  &  (\degr)	      & 		 & (Jy/beam)		     &  				      &  (Jy)                    &  (\arcsec)                &  (\arcsec)                 &  (\degr)          \\	       
  (35)           &   (36) ~ $\pm$ ~ (37)     &  (38)                                  &  (39)                         &    (40) ~ $\pm$ ~ (41)    &  (42)                     &  (43)                     &  (44)             &  (45)            &    (46) ~ $\pm$ ~ (47)    &  (48)                                  &    (49) ~ $\pm$ ~ (50)   &  (51)                     &  (52)                      &  (53)             \\	
\toprule[0.8pt] 
$\cdots$         &                           &                                        &                               & 			  &			      & 			  &		      & 		 &			     &  				      &                          &                           &                            &                   \\
1285.0	 & 7.16e+01 ~ 5.1e+00        &   2.57  			              &  7.10e+01		      &  7.79e+01 ~ 5.1e+00       &  25			      &  25			  &  162	      &    947.0         &  4.33e+01 ~ 1.8e+00       &  2.55                                  &  4.25e+01 ~ 1.8e+00      &  36			     &  36			  &  177  	      \\
$\cdots$         &                           &                                        &                               & 			  &			      & 			  &		      & 		 &			     &  				      &                          &                           &                            &                   \\
3070.0	 & 1.37e+02 ~ 5.2e+00        &   3.54  			              &  1.64e+02		      &  2.21e+02 ~ 5.3e+00       &  38			      &  25			  &   55	      &    2323.0	 &  9.41e+01 ~ 2.1e+00       &   4.88                                  &  1.32e+02 ~ 2.1e+00      &  46			     &  36			  &    48  	      \\
$\cdots$         &                           &                                        &                               & 			  &			      & 			  &		      & 		 &			     &  				      &                          &                           &                            &                   \\
9581.0	 & 2.77e+02 ~ 3.3e+00        &   9.83  			              &  2.75e+02		      &  2.68e+02 ~ 3.3e+00       &  25			      &  25			  &  68	          &   4397.0	 &  1.33e+02 ~ 1.3e+00       &  11.70                                  &  1.25e+02 ~ 1.3e+00      &  36			     &  36			  &  208  	      \\
$\cdots$         &                           &                                        &                               & 			  &			      & 			  &		      & 		 &			     &  				      &                          &                           &                            &                   \\
\midrule[0.5pt]
\end{tabular}
}
\scalebox{1.2}{$\sim$}
\vspace{0.2cm}

\scalebox{1.2}{$\sim$}
{\renewcommand{\arraystretch}{0.5}
\begin{tabular}{c c c c c c c c c c c c|}  
\toprule[1.0pt]\toprule[0.5pt] 
Sig$_{\rm N_{H_2}}$& $N^{\rm peak}_{\rm H_2}$ &  $N^{\rm peak}_{\rm H_2}$/$N_{\rm bg}$ &  $N^{\rm conv,500}_{\rm H_2}$ &  $N^{\rm bg}_{\rm H_2}$  &  FWHM$^{\rm a}_{\rm N_{H_2}}$ &  FWHM$^{\rm b}_{\rm N_{H_2}}$ & PA$_{\rm N_{H_2}}$  &  N$_{\rm SED}$    &  Core type     & SIMBAD                           & Comments    \\
                   & (10$^{21}$ cm$^{-2}$)    &                                        & (10$^{21}$ cm$^{-2}$)         &  (10$^{21}$ cm$^{-2}$)   &  (\arcsec)  	          &  (\arcsec)                    &  (\degr)            &                 &        &                &                                     \\ 
 (54)              & (55)                     &  (56)                                  & (57)                          &  (58)                    &  (59)                         &  (60)                         &  (61)               &    (62)         &  (63)  & (64)           & (65)                                              \\
\toprule[0.8pt] 																									            
$\cdots$           &                          &                                        &                               &                          &                               &                               &                     &                 &        &                &                                                              \\
1469.0              & 140.1		      &   3.31  			       &    44.2		               &  42.4	                  &  24			          &  19			          &  98		&		4       & protostellar       &       VLA 1623-243                                     &             \\
$\cdots$           &                          &                                        &                               &                          &                               &                               &                     &                 &        &                &                                                              \\
2691.0              & 247.0		      &   4.12  			       &    92.2		               &  59.9	                  &  35			          &  18			          &  53		&		4       & prestellar     &         LFAM 6                               &     SM1        \\
$\cdots$           &                          &                                        &                               &                          &                               &                               &                     &                 &        &                &                                                              \\
6877.0              & 497.4		      &   9.59  			       &   125.0		               &  51.9	                  &  18			          &  18			          &  62	        &		4   & protostellar   & IRAS 16293-2422A              &             \\
$\cdots$           &                          &                                        &                               &                          &                               &                               &                     &                 &        &                &                                                              \\
\midrule[0.5pt]
\end{tabular}
}
\tablefoot{Catalog entries are as follows: 
{\bf(1)} Core running number;
{\bf(2)} Core name $=$ HGBS\_J prefix directly followed by a tag created from the J2000 sexagesimal coordinates; 
{\bf(3)} and {\bf(4)}: Right ascension and declination of core center; 
{\bf(5)}, {\bf(15)}, {\bf(25)}, {\bf(35)}, and {\bf(45)}: Detection significance from monochromatic single scales, in the 70, 160, 250, 350, and 500~$\mu$m maps, respectively. 
(NB: the detection significance has the special value of $0.0$ when the core is not visible in clean single scales); 
{\bf(6)}$\pm${\bf(7)}, {\bf(16)}$\pm${\bf(17)} {\bf(26)}$\pm${\bf(27)} {\bf(36)}$\pm${\bf(37)} {\bf(46)}$\pm${\bf(47)}: Peak flux density and its error in Jy/beam as estimated by \textsl{getsources};
{\bf(8)}, {\bf(18)}, {\bf(28)}, {\bf(38)}, {\bf(48)}: Contrast over the local background, defined as the ratio of the background-subtracted peak intensity to the local background intensity ($S^{\rm peak}_{\rm \lambda}$/$S_{\rm bg}$); 
{\bf(9)}, {\bf(19)}, {\bf(29)}, {\bf(39)}: Peak flux density measured after smoothing to a 36.3$\arcsec$ beam; 
{\bf(10)}$\pm${\bf(11)}, {\bf(20)}$\pm${\bf(21)}, {\bf(30)}$\pm${\bf(31)}, {\bf(40)}$\pm${\bf(41)}, {\bf(49)}$\pm${\bf(50)}: Integrated flux density and its error in Jy as estimated by \textsl{getsources}; 
{\bf(12)}--{\bf(13)}, {\bf(22)}--{\bf(23)}, {\bf(32)}--{\bf(33)}, {\bf(42)}--{\bf(43)}, {\bf(51)}--{\bf(52)}: Major \& minor FWHM diameters of the core (in arcsec), respectively, 
as estimated by \textsl{getsources}. (NB: the special value of $-1$ means that no size measurement was possible); 
{\bf(14)}, {\bf(24)}, {\bf(34)}, {\bf(44)}, {\bf(53)}: Position angle of the core major axis, measured east of north, in degrees; 
{\bf(54)} Detection significance in the high-resolution column density image;  
{\bf(55)} Peak H$_{2}$ column density in units of $10^{21}$ cm$^{-2}$ as estimated by \textsl{getsources} in the high-resolution column density image; 
{\bf(56)} Column density contrast over the local background, as estimated by \textsl{getsources} in the high-resolution column density image;
{\bf(57)} Peak column density measured in a 36.3$\arcsec$ beam; 
{\bf(58)} Local background H$_{2}$ column density as estimated by \textsl{getsources} in the high-resolution column density image; 
{\bf(59)}--{\bf(60)}--{\bf(61)}: Major \& minor FWHM diameters of the core, and position angle of the major axis, respectively, as measured in the high-resolution column density image; 
{\bf(62)} Number of {\it Herschel} bands in which the core is significant (Sig$_{\rm \lambda} >$ 5) and has a positive flux density, excluding the column density plane; 
{\bf(63)} Core type: starless, prestellar, or protostellar; 
{\bf(64)} Closest counterpart found in SIMBAD, if any, up to 6$\arcsec$ from the {\it Herschel} peak position;
{\bf(65)} Comments. }
\end{sidewaystable*}

\clearpage

\begin{sidewaystable*}[htbp]\tiny\setlength{\tabcolsep}{5.2pt}
\caption{Derived properties of the dense cores identified in the HGBS maps of the Ophiuchus region (template, full table only provided online).}
\label{tab_der_cat_cores}
{\renewcommand{\arraystretch}{0.9}
\begin{tabular}{|r|c|c c c c c c c c c c c c|} 
\toprule[1.0pt]\toprule[0.5pt]  
 rNO     & Core name         &  RA$_{\rm 2000}$ &  Dec$_{\rm 2000}$        & $R_{\rm core}$        &  $M_{\rm core}$   &  $T_{\rm dust}$   &  $N^{\rm peak}_{\rm H_2}$ &  $N^{\rm ave}_{\rm H_2}$   &   $n^{\rm peak}_{\rm H_2}$  &  $n^{\rm ave}_{\rm H_2}$   &  $\alpha_{\rm BE}$  & Core type     &  Comments \\
         & HGBS\_J*          &  (h m s)         &  (\degr~\arcmin~\arcsec) & (pc)                  &  ($M_\odot$)      &  (K)              &  (10$^{21}$ cm$^{-2}$)    &  (10$^{21}$ cm$^{-2}$)     &   (10$^{4}$ cm$^{-3}$)      &  (10$^{4}$ cm$^{-3}$)      &                     &		         &		     \\   
 (1)     & (2)               &  (3)             &   (4)                    & (5) ~~~~~~ (6)        &  (7) $\pm$ (8)    &  (9) $\pm$ (10)   &  (11)                     &  (12) ~~~~ (13)            &   (14)                      &  (15) ~~~ (16)             &         (17)        &    (18)       &  (19)     \\   
\toprule[0.8pt] 																				 
$\cdots$ &                   &                  &                          &                       &                   &                   &                           &                            &	                          &                            &                     &               &		     \\ 
  89     & 162626.6-242431   & 16:26:26.68      &  --24:24:31.0            &  8.1e-03 ~~~ 1.4e-02  &  2.234 ~~~ 0.318  &  13.6 ~~~ 0.6     &  190.64                   &  127.12 ~~~  402.69        &    225.87                   &  192.86 ~~~ 1087.41        &       0.07          &  protostellar &      VLA1623     \\
$\cdots$ &                   &                  &                          &                       &                   &                   &                           &                            &	                          &                            &                     &               &		     \\ 
  91     & 162627.6-242359   & 16:26:27.65      &  --24:23:59.3            &  1.1e-02 ~~~ 1.7e-02  &  7.868 ~~~ 1.062  &  12.7 ~~~~ 0.6    &  378.77                   &  322.26 ~~~  753.76        &    423.33                   &  414.80 ~~~ 1483.83        &       0.03          &  prestellar   &	SM1         \\
$\cdots$ &                   &                  &                          &                       &                   &                   &                           &                            &	                          &                            &                     &               &		     \\ 
 458     & 163223.1-242836   & 16:32:23.16      &  --24:28:36.5            &  6.1e-03 ~~~ 1.2e-02  &  4.659 ~~~ 0.910  &  16.6 ~~~ 1.6     &  302.91                   &  368.05 ~~~ 1472.20        &    375.98                   &  657.90 ~~~ 5263.18        &       0.03          &  protostellar     & IRAS~16293	         \\
$\cdots$ &                   &                  &                          &                       &                   &                   &                           &                            &	                          &                            &                     &               &		     \\ 
\midrule[0.5pt]
\end{tabular}
}
\tablefoot{Table entries are as follows: {\bf(1)} Core running number; {\bf(2)} Core name $=$ HGBS\_J prefix directly followed by a tag created from the J2000 sexagesimal coordinates; 
{\bf(3)} and {\bf(4)}: Right ascension and declination of core center; 
{\bf(5)} and {\bf(6)}: Geometrical average between the major and minor FWHM sizes of the core (in pc), as measured in the high-resolution column density map 
after deconvolution from the 18.2$\arcsec$ HPBW resolution of the map and before deconvolution, respectively.
(NB: Both values provide estimates of the object's outer {\it radius} when the core can be approximately described by a Gaussian distribution, as is the case 
for a critical Bonnor-Ebert spheroid); 
{\bf(7)} Estimated core mass ($M_\odot$) assuming the dust opacity law advocated by \citet{2014A&A...562A.138R}; 
{\bf(9)} SED dust temperature (K); {\bf(8)} \& {\bf(10)} Statistical errors on the mass and temperature, respectively, including calibration uncertainties, but excluding dust opacity uncertainties; 
{\bf(11)} Peak H$_2$ column density, at the resolution of the 500$~\mu$m data, derived from a graybody SED fit to the core peak flux densities measured in a common 36.3$\arcsec$ beam at all wavelengths; 
{\bf(12)} Average column density, calculated as $N^{\rm ave}_{\rm H_2} = \frac{M_{\rm core}}{\pi R_{\rm core}^2} \frac{1}{\mu m_{\rm H}}$, 
          where $M_{\rm core}$ is the estimated core mass (col. {\bf 7}), $R_{\rm core}$ the estimated core radius prior to deconvolution (col. {\bf 6}), and $\mu = 2.8$;
{\bf(13)} Average column density calculated in the same way as for col. {\bf 12} but using the deconvolved core radius (col. {\bf 5}) instead of the core radius measured prior to deconvolution;  
{\bf(14)} Beam-averaged peak volume density at the resolution of the 500~$\mu$m data, derived from the peak column density (col. {\bf 11}) assuming a Gaussian spherical distribution: 
          $n^{\rm peak}_{\rm H_2} = \sqrt{\frac{4 \ln2}{\pi}} \frac{N^{\rm peak}_{\rm H_2}}{\overline{FWHM}_{\rm 500}}$; 
{\bf(15)} Average volume density, calculated as
          $n^{\rm ave}_{\rm H_2} = \frac{M_{\rm core}}{4/3 \pi R_{\rm core}^3} \frac{1}{\mu m_{\rm H}}$, using the estimated core radius prior to deconvolution; 
{\bf(16)} Average volume density, calculated in the same way as for col. {\bf 15} but using the deconvolved core radius (col. {\bf 5}) instead of the core radius measured prior to deconvolution; 
{\bf(17)} Bonnor-Ebert mass ratio: $\alpha_{\rm BE} = M_{\rm BE,crit} / M_{\rm obs} $ (see text for details); 
{\bf(18)} Core type: starless, prestellar, or protostellar; 
{\bf(19)} Comments may be \textit{no SED fit} or \textit{tentative bound} (see text for details). 
}

\end{sidewaystable*}

\clearpage
\newpage

\section{Completeness of the prestellar core sample}

\begin{figure}[!h]
  \centering
  \includegraphics[width=\hsize]{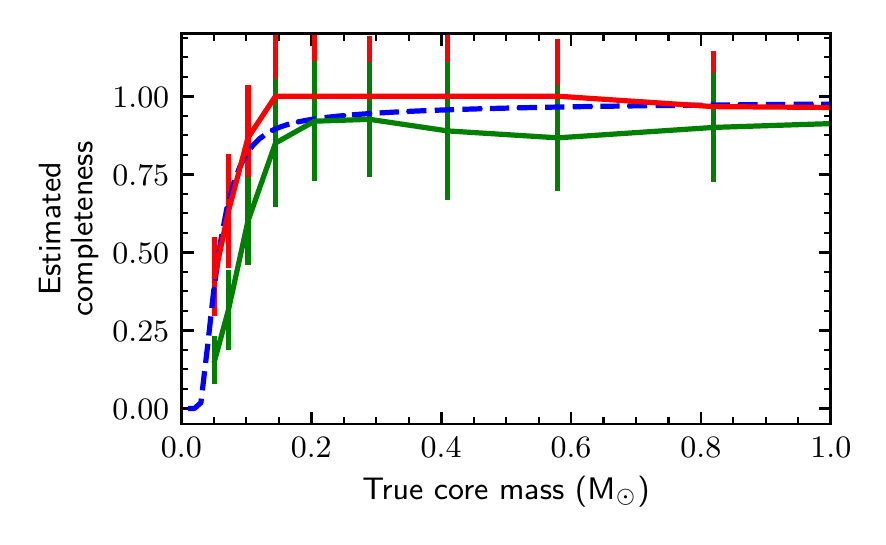}
  \caption{Completeness curves for our \textit{Herschel} census of candidate prestellar cores based on Monte-Carlo simulations using a population of synthetic prestellar cores detected by \textit{getsources} and classified as candidate prestellar cores (green solid curve) or classified as (bound or unbound) starless cores (red solid curve). 
For comparison, the dashed blue curve shows the completeness curve predicted by the simple model of the core extraction process described in Appendix B.2 of \cite{2015A&A...584A..91K}, scaled to the distance of the Ophiuchus cloud ($d= 139\,$pc).}
  \label{fig:complete}
\end{figure}

\begin{figure}[!h]
  \centering
  \includegraphics[width=\hsize]{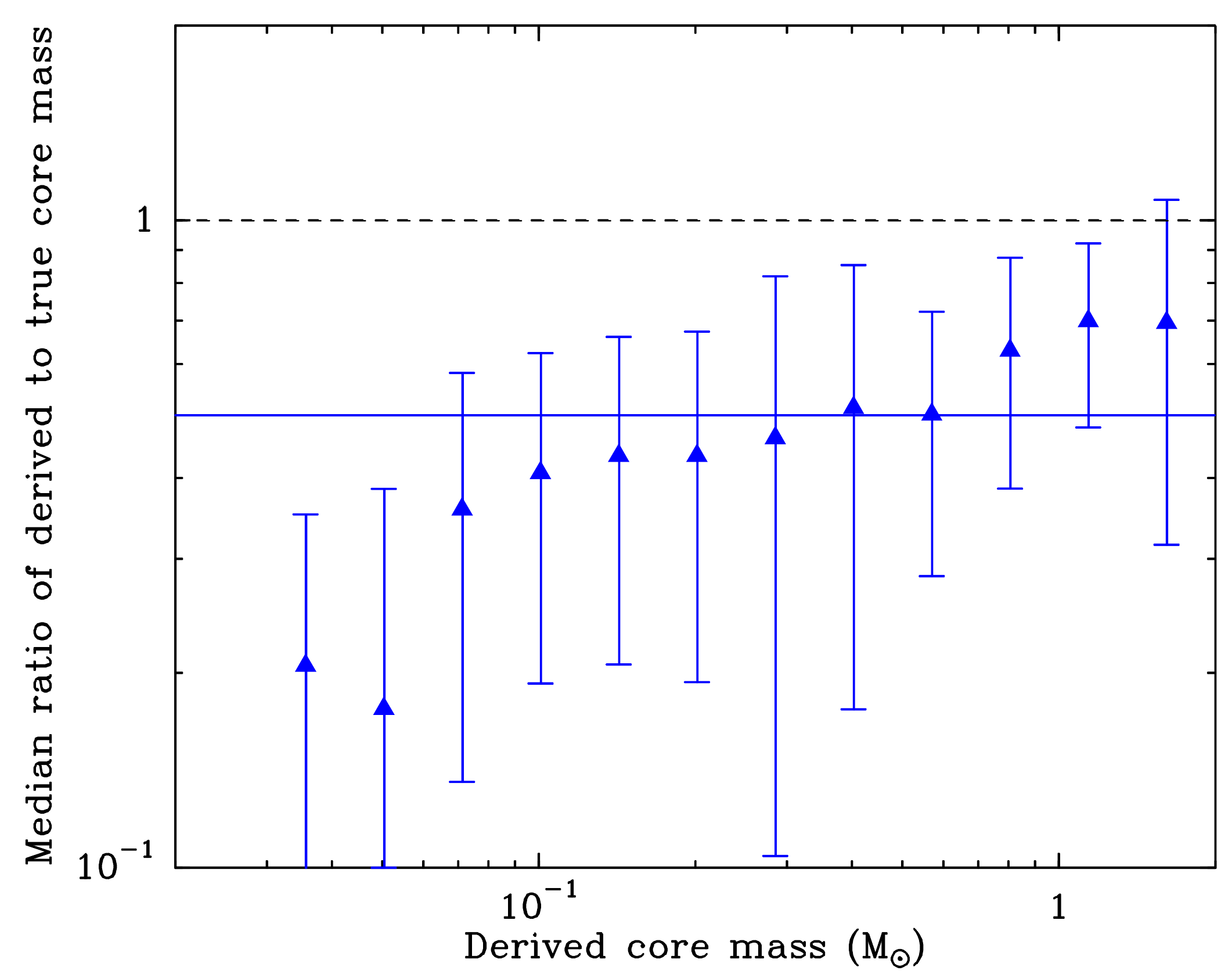}
  \includegraphics[width=\hsize]{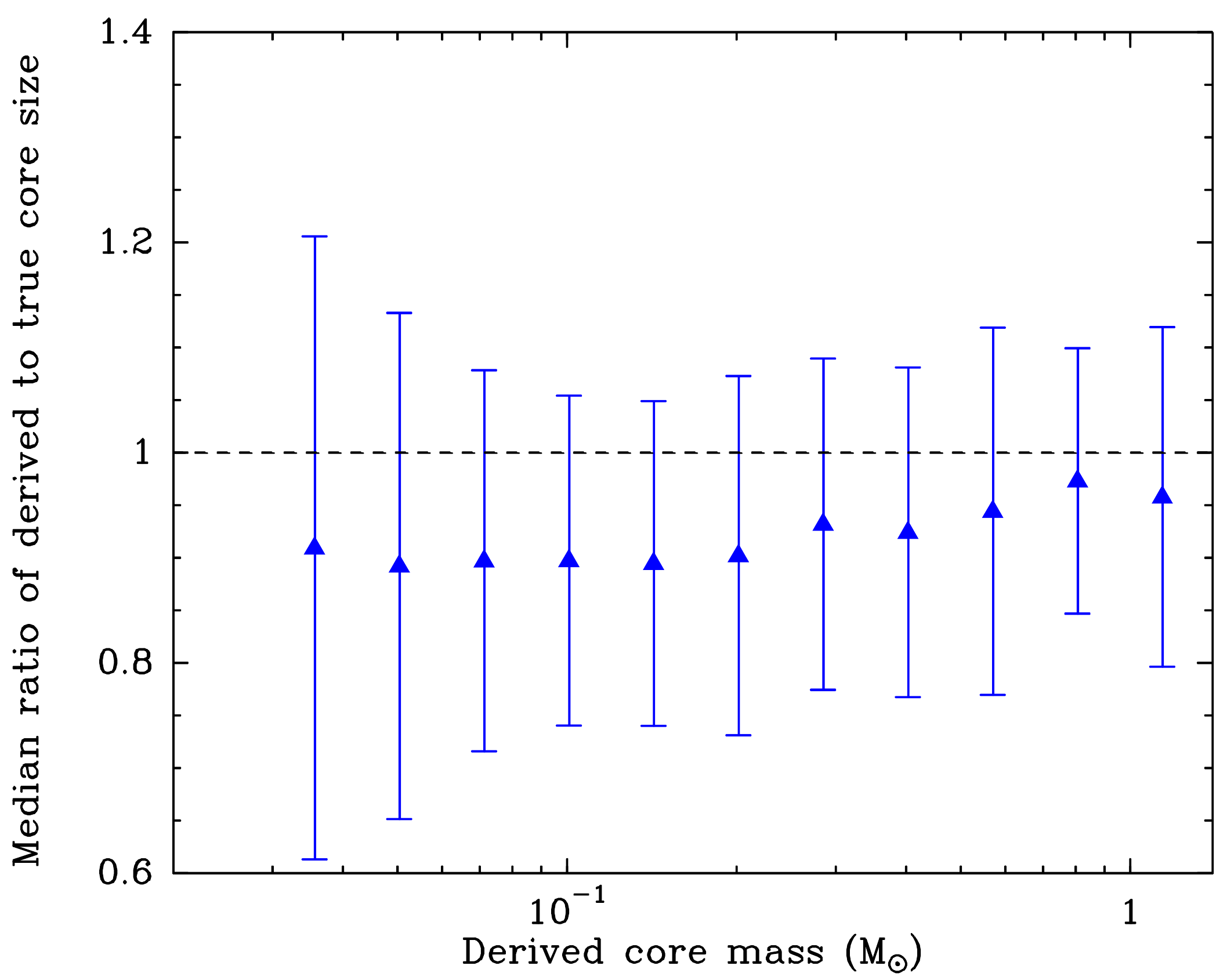}
  \includegraphics[width=\hsize]{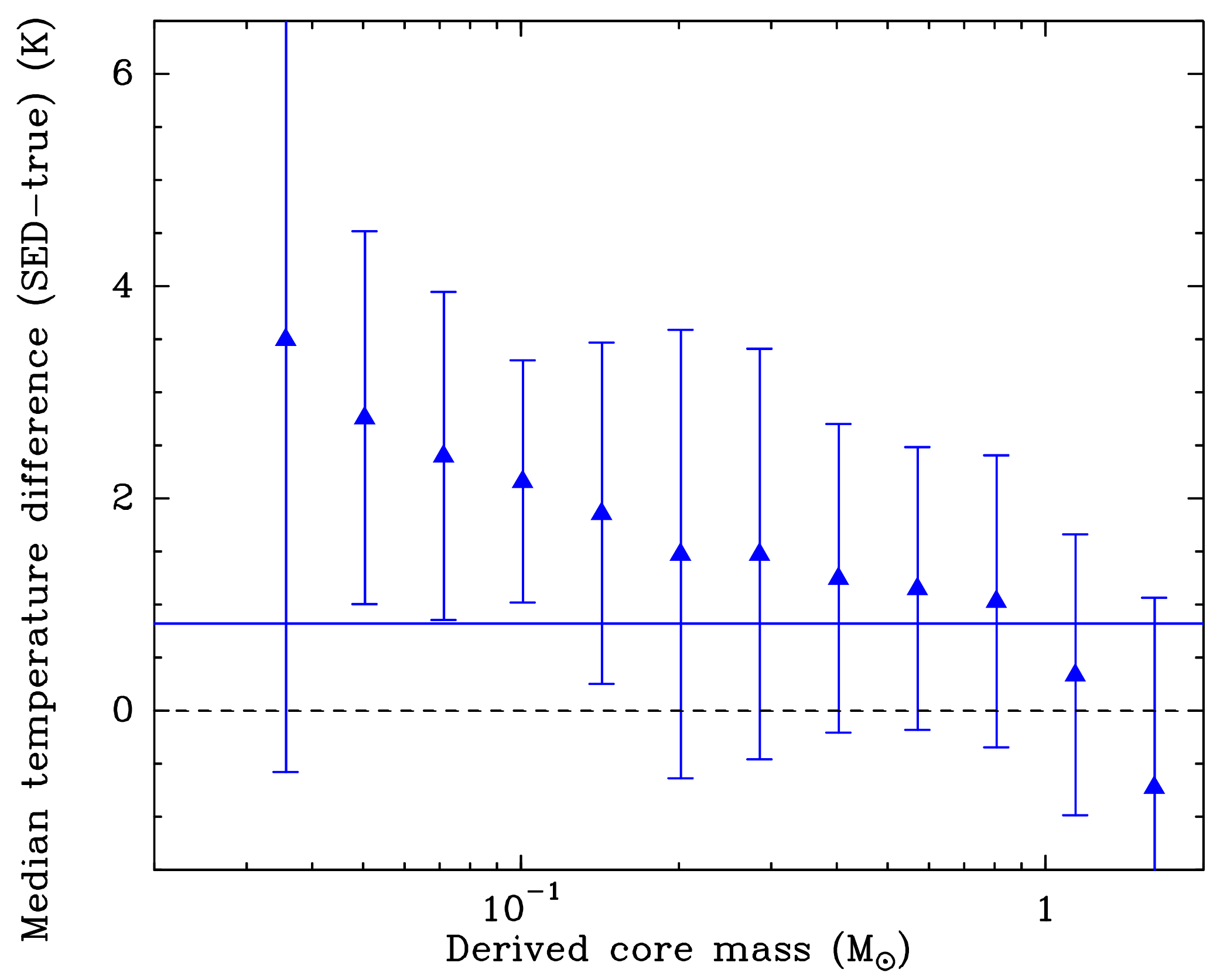}
  \caption{Comparison between the derived and the true values of the core masses (top panel), sizes (middle panel), and dust temperatures (bottom panel) in the Monte-Carlo simulations described here and in Sect.~\ref{sec:completeness}.}
  \label{fig:complete2}
\end{figure}

To estimate the completeness of the present {\it Herschel} census of prestellar cores in the Ophiuchus complex, we used both Monte-Carlo simulations (cf. Sect.~\ref{sec:completeness}) and the same model of the core identification process as described in Appendix~B.2 of \citet{2015A&A...584A..91K}. 
The same extraction steps with \textsl{getsources} and post-extraction selection criteria as described in Sect.~4.4 for the real {\it Herschel} data were applied to the simulated images. 
The resulting catalog of identified sources was then compared with the ``truth table'' of input synthetic cores injected in the simulated images to assess the completeness level of the  {\it Herschel} survey. 
Figure~\ref{fig:complete}) plots the fraction of successfully identified synthetic cores as a function true core mass.
As can be seen in this figure, our census of prestellar cores appears to be $>\,$$80\%$  complete above a true core mass of $\sim$0.1 \msun. 
In addition, the properties found for the cores identified in the simulated data were also compared to the true properties of the injected synthetic cores to estimate the accuracy of the main derived parameters (e.g., core mass, radius, temperature) given in Table~A.1 for the real cores. 
Figure~\ref{fig:complete2} (top panel) suggests that the derived core masses typically underestimate the true core masses by $\sim \,$50\% on average for observed masses between $\sim 0.1\, M_\odot $ and $\sim 1\, M_\odot $, around the peak of the observed prestellar CMF (Fig.~\ref{bound_unbound_cores_cmf}). 
This is most likely due to a slight  ($\sim 20$\%) underestimate of the core sizes (cf. middle panel of Fig.~\ref{fig:complete2}) and an overestimate of the mass-averaged dust temperatures by $\sim 1.5\,$K on average (cf. bottom panel of Fig.~\ref{fig:complete2}).

\end{document}